\documentclass{article}
\pdfoutput=1
\usepackage{enumerate}
\usepackage{amsfonts}
\usepackage{amsmath}
\usepackage{comment}
\usepackage{textcomp}
\usepackage{amsthm, tikz}
\usepackage{amssymb}
\usepackage{color}
\usepackage{graphicx}
\usepackage{bbm}
\usepackage[matrix,arrow,curve]{xy}
\usepackage{array}
\usepackage{mathtools}
\usepackage{hhline}

\def\be{\begin{eqnarray}}
\def\ee{\end{eqnarray}}
\def\nn{\nonumber}

\def\X{{\cal X}}
\def\Y{{\cal Y}}
\def\l[{\phantom.[}

\hoffset= - 1.4in         

\begin{document}

\title{{\bf {Colored HOMFLY polynomials for
the pretzel knots and links
}\vspace{.2cm}}
\author{{\bf A. Mironov$^{a,b,c,}$}\footnote{mironov@lpi.ru; mironov@itep.ru}, \ {\bf A. Morozov$^{b,c,}$}\thanks{morozov@itep.ru} \ and {\bf
A. Sleptsov$^{b,c,e,}$}\footnote{sleptsov@itep.ru}}
\date{ }
}

\maketitle

\vspace{-5.0cm}

\begin{center}
\hfill FIAN/TD-20/14\\
\hfill ITEP/TH-47/14\\
\end{center}

\vspace{4.2cm}

\begin{center}
$^a$ {\small {\it Lebedev Physics Institute, Moscow 119991, Russia}}\\
$^b$ {\small {\it ITEP, Moscow 117218, Russia}}\\
$^c$ {\small {\it National Research Nuclear University MEPhI, Moscow 115409, Russia }}\\
$^e$ {\small {\it Laboratory of Quantum Topology,
Chelyabinsk State University, Chelyabinsk 454001, Russia}}
\end{center}

\vspace{1cm}

\begin{abstract}
With the help of the evolution method we calculate all
HOMFLY polynomials in all symmetric representations $[r]$ for a huge family of (generalized) pretzel links, which are made from
$g+1$ two strand braids, parallel or antiparallel,
and depend on $g+1$ integer numbers.
We demonstrate that they possess a pronounced new structure: are decomposed into a sum of a product of $g+1$ elementary polynomials, which are obtained from the
evolution eigenvalues by rotation with the help of rescaled $SU_q(N)$ Racah matrix, for which we provide an explicit expression.
The generalized pretzel  family contains many mutants,  undistinguishable by
symmetric HOMFLY polynomials, hence, the extension of our results
to non-symmetric representations $R$ is a challenging open problem.
To this end, a non-trivial generalization of the suggested formula
can be conjectured for entire family with arbitrary $g$ and $R$.
\end{abstract}

\vspace{2cm}

\section{Introduction}

Despite impressive progress during the last years \cite{RT}-\cite{inds},
evaluation of colored HOMFLY polynomials \cite{pols} for particular knots and links
remains a non-trivial exercise.
It makes use of a variety of advanced methods of modern theoretical physics,
however, they still remain not powerful enough for this task, which
in turn helps to further develop these methods.
Knot polynomials are interesting, because they are the simplest
possible example of Wilson-loop averages in gauge (Chern-Simons) theory \cite{Wit}
on one hand and are close relatives of the holomorphic conformal
blocks on the other hand.
They depend on variety of parameters, and the purpose is
to study and understand these dependencies, which are already
known to satisfy various interesting equations, generalizing the
previously known ones in simpler (quantum) field theories,
\cite{MMeqs}.

Especially interesting are results obtained for entire
families of knots or links.
The most famous example is the two-parametric set of
{\it torus} knots and links, formed by a non-intersecting lines,
wrapping around a torus respectively $m$ and $n$ times along its two non-contractible cycles.
The link diagrams for a torus knot/link is just a closure
of especially simple $m$-strand braid,

\bigskip

\begin{figure}[h!]
\centering\leavevmode
\includegraphics[width=13 cm]{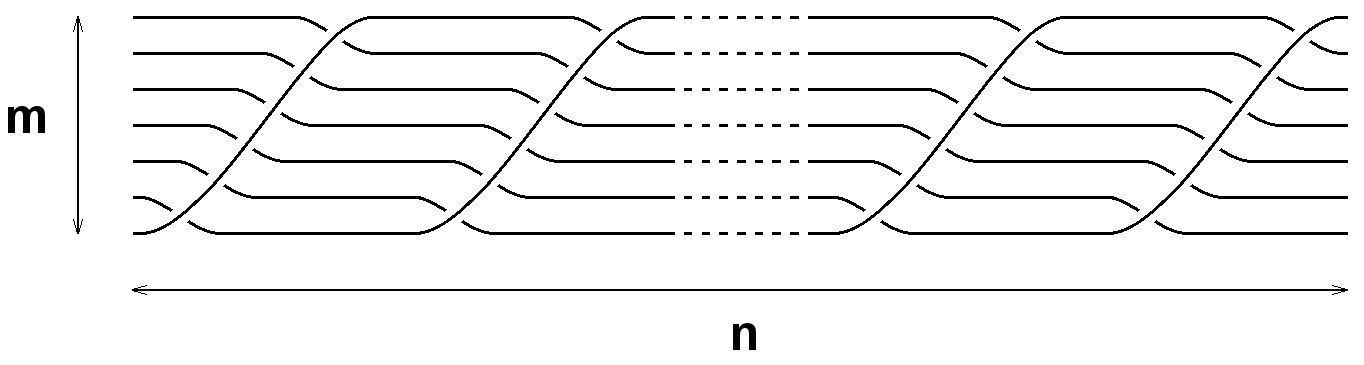}
\caption{Torus [m,n] braid}
\label{braidmn}
\end{figure}

In this case, the HOMFLY polynomial in arbitrary representation $R$ is
given by the Rosso-Jones formula \cite{RJ},
\be
H_R^{[m,n]}(q,A) =
q^{2mn\varkappa_R}A^{(m-1)n|R|}\cdot \sum_{Q\in R^{\otimes m}} C_{RQ} \lambda_Q^{2n/m}\chi_Q^*(q,A)
\label{RJ}
\ee
Here the sum goes over Young diagrams $Q$ of the size $|Q|=m|R|$,
the quantum dimensions of the corresponding representations of
the linear group $GL(N)$ are the values of Schur functions
at the "topological locus" in the space of time-variables,
$\chi_Q^*=\chi_Q\left\{p_k=\frac{\{A^k\}}{\{q^k\}}\right\}$
where $A=q^N$ and $\{x\}= x-x^{-1}$, so that the quantum number
is $[x]=\frac{\{q^x\}}{\{q\}}$ and the ``DGR differential" \cite{DGR}
is $D_i = \frac{\{Aq^i\}}{\{q\}}$.
Parameters $\lambda_Q$ are associated eigenvalues of the quantum
${\cal R}$-matrix, made from the eigenvalues $\varkappa_Q$ of
the cut-and-join operator \cite{MMN}
\be
\lambda_Q \sim q^{\varkappa_Q},\ \ \ \ \ \ \ \ \ \ \ \ \ \
\varkappa_Q = \sum_{(i,j)\in Q} (j-i)
\label{ladef}
\ee
where there is an arbitrary factor in $\lambda_Q$ that depends on the framing.
Finally, the coefficients $C_{RQ}$ are defined from the expansion of
the Adams transform of characters $\chi_R$:
\be
\Big(\chi_R\{p_{mk/l}\}\Big)^{l} =\sum_{Q\in R^{\otimes m}} C_{RQ}\chi_Q\{p_k\}
\ee
where $l = \text{maximal common divisor}(m,n)$ is the number of components
in the torus link.
For coprime $n$ and $m$ one has a knot and $l=1$.

When $A=q^N$, eq.(\ref{RJ}) can be also recast in an $N$-fold integral \cite{TBEM}
\be
H_R^{[m,n]}(q,A) \sim \int \prod_{i=1}^N d\mu_i \exp\left\{\frac{\mu_i^2}{2mn\hbar}\right\}
\prod_{i<j}^N \sinh\left(\frac{\mu_i-\mu_j}{2m}\right) \sinh\left(\frac{\mu_i-\mu_j}{2n}\right)
\, \chi_R\Big[\text{diag}(e^{\mu_i})\Big]
\label{TBEM}
\ee
where the symmetry between $m$ and $n$ is explicitly restored.

Eq.(\ref{RJ}) can be directly generalized to superpolynomials \cite{AgS,DMMSS,Che},
depending on one extra parameter $t$, with the Schur functions promoted to
the Macdonald polynomials, though associated deformation of the matrix model
(\ref{TBEM}) is still unavailable.

Also a mystery remains what makes the torus links so special: despite numerous
attempts no comparably explicit formulas for all representations $R$ at once
were yet found for any other family.
What was done, however, the simple dependence on $n$ (but not on $m$) in (\ref{RJ})
was interpreted in \cite{DMMSS} as an evolution in the length of an $m$-strand braid,
and this fact remains true for any such braid inside any, arbitrarily complicated
knot or link \cite{evo}: dependence on its length $n$ will enter only through a linear
combination of $\lambda_Q^{2n/m}$.
Still the coefficients $C_{RQ}\chi_Q^*$ can be quite sophisticated.
To define them, one needs "initial conditions" for the evolution, i.e.
explicit knowledge of knot polynomials for a few particular values of $n$.
Despite an extreme naiveness of the evolution method it allowed one to study
certain interesting families, in particular, the important family of twist knots
\cite{evo} and led to a discovery of a very important "differential structure"
\cite{IMMMfe,artdiff} of arbitrary knot polynomials, which seems related to
the original ideas in \cite{DGR}, and led to a number of impressive advances
in knot calculus, at least, for symmetric representations \cite{GS,FGS1,FGS2,Rama,Rama2,advdiff,GGS,AENV}.
(However, attempts to generalize the matrix model (\ref{TBEM}) in \cite{Almamo}
and to describe non-symmetric representations in \cite{Ano21,germ21,MMM21}
are still only partly successful.)

\begin{figure}[h!]
\centering\leavevmode
\includegraphics[width=18 cm]{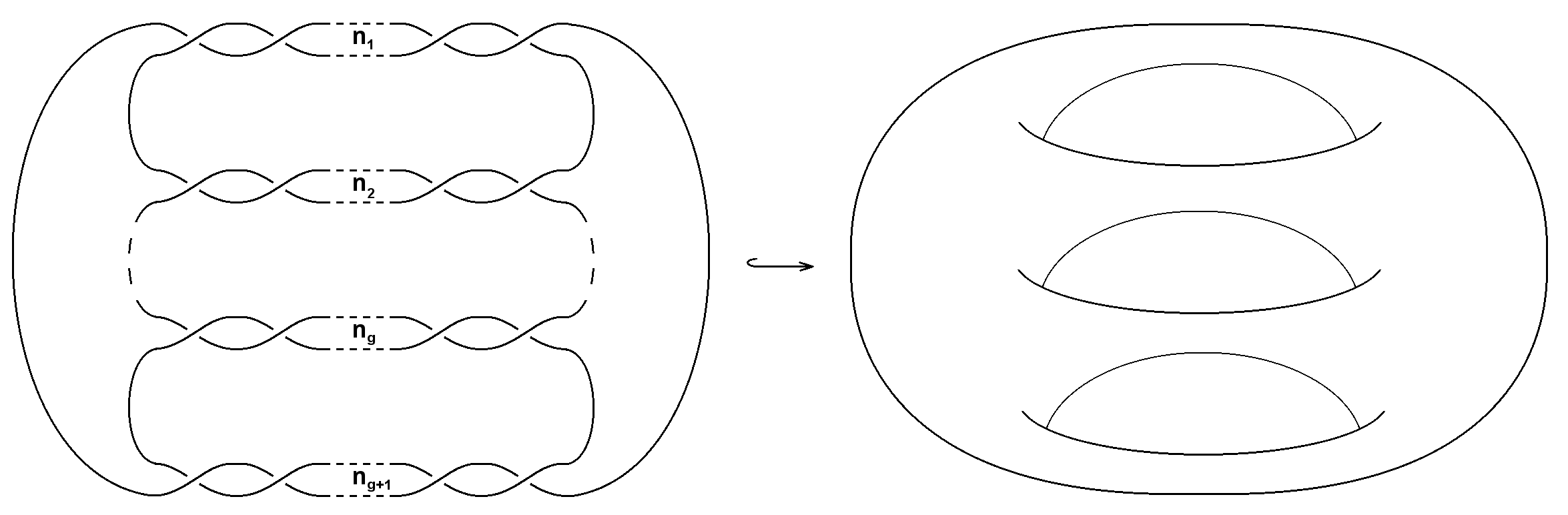}
\caption{Pretzel link or knot of genus $g=3$ }
\label{braidg}
\end{figure}

The goal of the present paper is to extend previous calculations to a much richer family,
which, taken as a total, looks like a straightforward generalization
of the torus knots, and thus provides more chances to guess the relevant way
to generalize (\ref{RJ}) and, perhaps, even (\ref{TBEM}).
These are knots and links formed by wrapping around a surface of genus $g$ without self-intersections,
which can be different from $g=1$.
The simplest set of this type has a link diagram (see Figure 2), consisting of $g+1$
two-strand braids, and thus has $g+1$ different evolution parameters
$n_1,\ldots, n_{g+1}$ (for $g=1$ everything depends on the sum $n=n_1+n_2$). In literature (see \cite{pretz}) this family is known as the pretzel knots and links.
The family is actually split into subfamilies, differing by mutual orientation
of strands in the braids.
For certain orientations the family has a cyclic symmetry $n_k\longrightarrow n_{k+1}$.
In fact, if one considers only symmetric representations, the symmetry
is actually enhanced to arbitrary permutations of $n_k$, links/knots
related by these permutations are actually mutants \cite{mut}
and symmetric HOMFLY polynomials are the same for them.

\paragraph{Notations.} For the sake of convenience, we repeat here our notations once again:
\be
\chi_Q^*=\chi_Q\left\{p_k=\frac{\{A^k\}}{\{q^k\}}\right\}\nn\\
\{x\}= x-x^{-1},\ \ \ \ \ [x]=\frac{\{q^x\}}{\{q\}},\ \ \ \ \ D_i = \frac{\{Aq^i\}}{\{q\}}\\
\lambda_Q \sim \epsilon_Qq^{\varkappa_Q},\ \ \ \ \ \
\varkappa_Q = \sum_{(i,j)\in Q} (i-j)\nn
\ee
where $\epsilon_Q$ is a sign factor, which will be fixed latter (it is always +1 in the Rosso-Jones case).
As soon as throughout the text only the Schur functions at the topological locus, $\chi_Q^*$ are used (for the only exception see the third paragraph of section \ref{ex}), from now on, we omit the asterisk and use just the notation $\chi_Q$.

\section{Warm-up examples}

\subsection{Genus $g=1$, fundamental representation, two parallel strands}

We begin with this simplest example,
which  is the simplest possible case of the Rosso-Jones formula (\ref{RJ}).
In our family we should restrict it to two strands, $m=2$, so that
\be
H^{(n_1,n_2)}_R = \sum_{Q\vdash 2|R|}  \lambda_Q^{n_1+n_2} \chi_Q
\ee
and, in the fundamental representation,
\be
H^{(n_1,n_2)}_{[1]} = \lambda_{[2]}^{n_1+n_2}\chi_{[2]} + \lambda_{[11]}^{n_1+n_2}\chi_{[11]}
\ee
with $\lambda_{[2]}=q/A$ and $\lambda_{[11]}=-1/(qA)$ in the topological framing.

\begin{figure}[h!]
\centering\leavevmode
\includegraphics[width=8 cm]{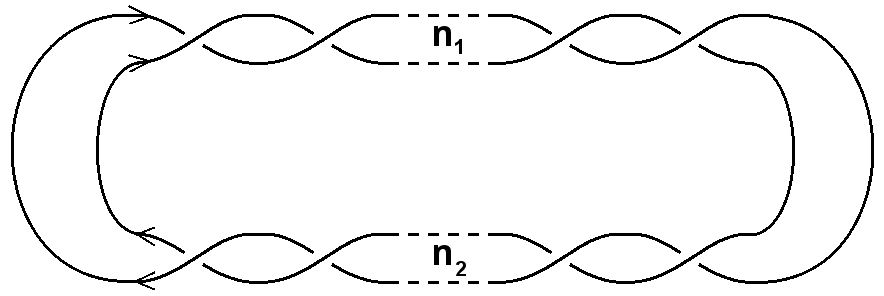}
\caption{Two parallel strands: torus links and knots}
\label{par_1}
\end{figure}

However, if one did not know the answer and looks at the problem
from the point of view of the evolution method, it is necessary to consider
the following anzatz:
\be
H^{(n_1,n_2)}_{[1]}=
c_{11}\lambda_{[2]}^{n_1+n_2}+ c_{10} \lambda_{[2]}^{n_1}\lambda_{[11]}^{n_2} +
c_{01}\lambda_{[2]}^{n_2}\lambda_{[11]}^{n_1}
+ c_{00}\lambda_{[11]}^{n_1+n_2}
\ee
with four unknown coefficients.
Apparent symmetry between $n_1$ and $n_2$ implies that $c_{10}=c_{01}$,
and looking at the picture one understands that the answer depends only on
$n_1+n_2$, thus actually $c_{10}=c_{01}=0$.
The two remaining parameters can be found from the two initial conditions:
for $n_1+n_2 = \pm 1$ one gets the unknot, with the HOMFLY polynomial equal to
$\chi_{[1]}$, i.e.
\be
c_{11}\lambda_{[2]}^{\pm 1} + c_{00}\lambda_{[11]}^{\pm1} = \chi_{[1]}
\ee
and
\be
c_{11} = \frac{\frac{1}{\lambda_{[11]}}-\lambda_{[11]}}{\frac{\lambda_{[2]}}
{\lambda_{[11]}}-\frac{\lambda_{[11]}}{\lambda_{[2]}}}\cdot \chi_{[1]} =
\frac{\{Aq\}}{\{q^2\}}\cdot \chi_{[1]} = \chi_{[2]}
 \nn \\
c_{00} = \frac{\lambda_{[2]}-\frac{1}{\lambda_{[2]}}}{\frac{\lambda_{[2]}}
{\lambda_{[11]}}-\frac{\lambda_{[11]}}{\lambda_{[2]}}}\cdot\chi_{[1]} =
 \frac{\{A/q\}}{\{q^2\}} \cdot \chi_{[1]} = \chi_{[11]}
\ee
what brings us back to
\be
H^{(n_1,n_2)}_{[1]} = \lambda_{[2]}^{n_0+n_1}\chi_{[2]} + \lambda_{[11]}^{n_0+n_1}\chi_{[11]}
\ee
(one can easily check the third obvious initial condition: for $n_0+n_1=0$
one gets a pair of disconnected unknots with the HOMFLY polynomial
$\chi_{[1]}^2 = \chi_{[2]}+\chi_{[11]}$).
Since $\chi_{[2]}+\chi_{[11]} =   \frac{\{Aq\}+\{A/q\}}{\{q^2\}}\cdot\chi_{[1]}=
\chi_{[1]}^2$
(the relation is actually valid beyond the topological locus),
one can rewrite this in an identical, but more sophisticated form:
\be
H^{(n_1,n_2)}_{[1]} =
\frac{\chi_{[2]}^2+\chi_{[2]}\chi_{[11]}}{\chi_{[1]}^2}\cdot\lambda_{[2]}^{n_1+n_2}
+ \frac{\chi_{[2]}\chi_{[11]}-\chi_{[2]}\chi_{[11]}}{\chi_{[1]}^2}
\Big(\lambda_{[2]}^{n_1}\lambda_{[11]}^{n_2}+\lambda_{[2]}^{n_2}\cdot\lambda_{[11]}^{n_1}\Big)
+ \frac{\chi_{[11]}^2+\chi_{[2]}\chi_{[11]}}{\chi_{[1]}^2}\lambda_{[11]}^{n_1+n_2}=\nn \\
= \sum_{i=0}^2 C^{\,i}_{[1]}\cdot
\Big(\lambda_{[2]}^{n_1+n_{i}}\lambda_{[11]}^{n_{i+1}+n_2}+
\text{permutations of}\ n_1, n_2\Big)
\label{expanH1}
\ee
where
\be
 C^{\,i}_{[1]} = \frac{1}{\chi_{[1]}^2}\Big(\chi_{[2]}^{2-i}\chi_{[11]}^{i}
+ (-)^i\chi_{[2]}\chi_{[11]}\Big)
\ee
and the only permutations from the two different groups of indices are included.

\subsection{Genus $g=1$, fundamental representation, antiparallel strands}

Before going to higher $g$ and higher representations, we consider the same genus-one two-strand example,
but now with antiparallel strands:

\begin{figure}[h!]
\centering\leavevmode
\includegraphics[width=8 cm]{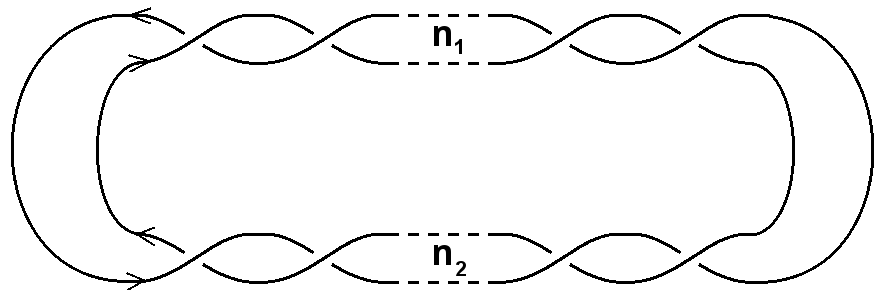}
\caption{Two antiparallel strands: torus links}
\label{anti_1}
\end{figure}

\noindent
This configuration is possible only if $n_1+n_2$ is even, and it is always a link,
hence generically the corresponding HOMFLY polynomials depend on two representations, $R_1\otimes R_2$.
The two parallel strands, considered in the previous section,
correspond to $R_2=R_1=[1]$, while for the antiparallel strands the fundamental HOMFLY implies that $R_2$ is rather
conjugate of $R_1$, $R_2=\overline{R_1}=\overline{[1]}= [q^{N-1}]$.
This is still a particular case of the Rosso-Jones formula (\ref{RJ}),
since it is valid at any representation.

From the point of view of the evolution method, one has now
$[1]\otimes\overline{[1]} = \text{Adjoint} + \text{singlet}$,
and according to \cite{evo} the two relevant eigenvalues are
$\lambda_{0} = 1$ and $\lambda_{adj}=-A$.
As the initial condition one can take the pair of unknots at $n_1+n_2=0$
and the Hopf link at $n_1+n_2=2$.

\subsection{HOMFLY in the fundamental representation at arbitrary genus \label{symgf}}

\begin{figure}[h!]
\centering\leavevmode
\includegraphics[width=16 cm]{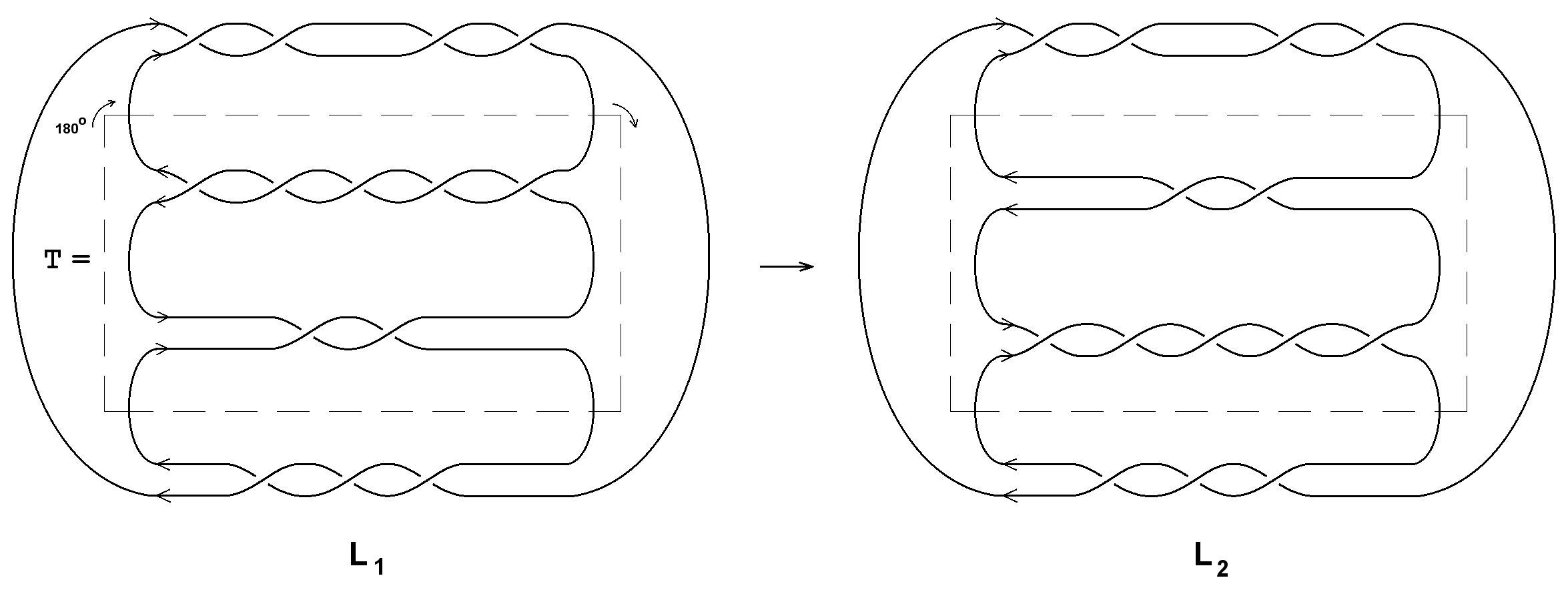}
\caption{Mutant knots}
\label{mutant}
\end{figure}

Now we can switch to arbitrary genus.
Again there will be different options to choose orientations of particular strands.
While for links the freedom is rather big, for knots the orientation depends only
on the genus.
For odd $g$ one can make all the braids parallel,
while for even $g$ exactly one should be antiparallel. Moreover, the
corresponding parameter, which we choose to be $n_{g/2+1}$, should be even.
 We also consider the case when all the braids are antiparallel.

The next question is what happens to the symmetry $n_1\leftrightarrow n_2$.
For $g>1$ the polynomials depend on all $n_i$ independently,
and there is only a cyclic symmetry when all $n_k\longrightarrow n_{k+1}$.
However, as we shall see, the answer in the fundamental representation is
actually symmetric in all $n_k$. In fact this should not come as a surprise, because permutation of the two adjacent $n_k$'s is just a knot mutation. Indeed, (by definition following \cite{crom}) let we have oriented link $L_1$ which contains a marked tangle $T$ (see Figure \ref{mutant}). Remove $T$, rotate it by $180^{\circ}$ about the axe transversal to the plane of the picture and glue it back in position to form a new link $L_2$. If $L_1 \neq L_2$ then they are called mutants of each other, and this operation is called mutation. Since the HOMFLY polynomials in symmetric representations do not distinguish the mutant knots \cite{mut}, with the help of mutation one can permute $n_k \leftrightarrow n_{k+1}$.
This enhanced symmetry reduces the number of necessary initial conditions and, thus, more formulas can be obtained and more are the chances to observe regularities, leading to discovery of generic expressions. Many are provided, by putting one of parameters, say $n_g$, equal to zero,
then the knot/link reduces to a composite one, which enjoys the
decomposition property
\be
\frac{H^{\text{composite}}_{R}}{\chi_R} =
\prod_{\text{components}} \frac{H^{\text{component}}_R}{\chi_R}
\ee
To these patterns, one can add already known particular examples, like
twist knots.

All this makes explicit calculation by the evolution method
possible at genera $g=1,2,3,4$, at least in the fundamental representation.
And this is enough to discover the structure and obtain the general formulas
for the HOMFLY polynomial in the fundamental representation:
\be
H^{(n_1,\ldots,n_{g+1})}_{[1]} =
\sum_{i=0}^{g+1} C^{\,i}_{[1]}\cdot
\Big(\lambda_{0}^{n_1+\ldots+n_{i}}\lambda_{1}^{n_{i+1}+\dots+n_{g+1}}+
\text{permutations of}\ n_i\Big)
\label{Hf}
\ee
where again only permutations from the two different groups of indices are included and the coefficients $C^{\,i}_{[1]}$ are:

\noindent

$\bullet$ odd $g$, all braids parallel:
\be
C^i_{[1]} =\frac{1}{\chi_{[1]}^{g+1}} \Big(\chi_{[2]}^i\chi_{[11]}^{g+1-i}
+ (-)^i \chi_{[2]}\chi_{[11]} z^{g-1}\Big)\\
\lambda_{0}=\lambda_{[11]},\ \lambda_{1}=\lambda_{[2]}
\label{Hfpar}
\ee

$\bullet$ even $g$, all braids antiparallel:
\be
C^{\bar i}_{[1]} =
\frac{1}{z^i \chi_{[1]}^{g+1}}\Big(\chi_{[2]}^i\chi_{[11]}  + (-)^i\chi_{[2]}\chi_{[11]}^i \Big)\\
\lambda_{0}=1,\ \lambda_{1}=\lambda_{adj}
\label{Hfantipar}
\ee

$\bullet$ even $g$, all braids parallel, except for one antiparallel, $n_{g/2+1}$ should be even; in this case each term in the sum,
(\ref{Hf}) is a product of $g$ factors $\lambda_{[11]}^{n_i}$ and $\lambda_{[2]}^{n_j}$ and one factor either $\lambda_{0}^{n_k}=1$ or
$\lambda_{adj}^{n_k}$, in these two different cases the coefficients $C^{i}_{[1]}$ being
\be
e.v.\ \lambda_{0}=1: &
C^{i}_{[1]} =\frac{1}{\chi_{[1]}^{g+1}} \Big(\chi_{[2]}^i\chi_{[11]}^{g-i}
+ (-)^i \chi_{[2]}\chi_{[11]} z^{g-2}\Big) \nn \\
e.v.\ \lambda_{adj}=A: &
C^{i}_{[1]} =\frac{\chi_{[2]}\chi_{[11]}}{z^2\chi_{[1]}^{g+1}} \Big(\chi_{[2]}^i\chi_{[11]}^{g-i}
+ (-)^{i+1}  z^{g}\Big) =
\frac{\{q\}^{g-1}\{Aq\}\{A/q\}}{\{A\}^{g+1}}\Big(\chi_{[2]}^i\chi_{[11]}^{g-i}
+ (-)^{i+1}  z^{g}\Big)
\label{gen2pap}
\ee
Here $z= \frac{1}{[2]}\chi_{[1]}$.

The  common factor $\{Aq\}\{A/q\}$ in the second formula in (\ref{gen2pap})
is required by the differential expansion.

\bigskip

The structure of these formulas is very simple:
there is the "main contribution", the first terms in each line,
which is in a clear one-to-one correspondence with the combination of $\lambda$-factors,
plus "corrections" which look a little less universal.
In fact, the same structure survives in higher representations,
at least symmetric.

\bigskip

{\bf Formulas (\ref{Hfpar})-(\ref{gen2pap})
provide an exhaustive description of the fundamental HOMFLY
for {\it all} the pretzel knots.}

\bigskip

For $N=2$ there is no orientation dependence (except for a simple framing
factor\footnote{This is because the orientation
independence is due to a group theory argument:
for $SU(2)$ group the representation coincides with its conjugate.
However, this is only the vertical framing that respects the group theory structures,
and in the topological framing there is slight orientation dependence, that is, additional factors.}),
and all the four formulas turn into one:
\be
\boxed{
C_{[1]}^{(i)} \ \stackrel{A=q^2}{=}\ \frac{[3]}{[2]^{2g+2}}\Big([3]^{i-1}+(-)^i\Big)
}
\label{fJones}
\ee
what gives $1,\ 0,\ [3],\ \frac{[3][4]}{[2]}=[5]+1,\ \ldots$  for $i=0,1,2,3,\ldots$ respectively.
(Note that at $N=2$ we have $z=1$, $\chi_{[2]}=[3]$ and $\chi_{[11]}=1$.)
Eq.(\ref{fJones}) is in perfect accordance with the result in \cite{GMMMS,GalS},
as well as with those in \cite{PretzJones}.

\section{Main result: arbitrary symmetric representation $[r]$}

Our main result is an explicit combinatorial formula for unreduced HOMFLY of arbitrary pretzel link in symmetric representation. The formula includes only {\it three} ingredients and looks like
\be
 \boxed{ \ \
H^{n_1,\ldots,n_{g+1}}_R \ \ \ \  \ = \ \ \ \ \
\sum_{X} \dim_qX \, \prod_{i=1}^{g+1} \, \sum_Y {\cal A}_{XY}\lambda_Y^{n_i}
}
\label{poldeco1}
\ee
Now we define the ingredients.

\bigskip

$\bullet \ {\bf Eigenvalues.}$ Since we construct the pretzel link with the help of 2-strand braids only, there are only two possible orientations for such a braid: \
\begin{tikzpicture}[scale=0.5]
\draw[line width=1,white]
(1,1) -- (1,1);
\draw[line width=1]
(0,0) -- (1,0) -- (2,0.8) -- (3,0.8);
\draw[line width=1]
(0,0.8) -- (1,0.8);
\draw[line width=1]
(1,0.8) -- (1.375,0.5) (1.624,0.3)  -- (2,0) -- (3,0);
\draw[line width=1,->,-stealth] (0,0) -- (0.75,0);
\draw[line width=1,->,-stealth] (0,0.8) -- (0.75,0.8);
\end{tikzpicture}
 \ parallel  and  \
\begin{tikzpicture}[scale=0.5]
\draw[line width=1,white]
(1,1) -- (1,1);
\draw[line width=1]
(0,0) -- (1,0) -- (2,0.8) -- (3,0.8);
\draw[line width=1]
(0,0.8) -- (1,0.8);
\draw[line width=1]
(1,0.8) -- (1.375,0.5) (1.624,0.3)  -- (2,0) -- (3,0);
\draw[line width=1,->,-stealth] (0,0) -- (0.75,0);
\draw[line width=1,->,-stealth] (1,0.8) -- (0.25,0.8);
\end{tikzpicture}  \ antiparallel. The parallel strands correspond to the product of two  symmetric representations $[r]$:
\be
\label{parrep}
\l[r]\otimes [r] = \oplus_{m=0}^r\ [\,r+m,\,r-m]
\ee
The corresponding evolution eigenvalues $\lambda$ in the topological framing are equal to
\be
\lambda_m = (-)^{m+1} \, \frac{q^{\varkappa_{[r+m,r-m]}}}{A^r\cdot q^{4\varkappa_{[r]}}}=(-)^{m+1}{q^{-r^2+m^2+m}\over A^r}
\label{lamdef}
\ee
Similarly, the antiparallel strands correspond to the product of symmetric representation and its conjugate:
\be
\label{aprep}
\l[r]\otimes\overline{[r]} = \oplus_{m=0}^r \ [2m,m^{N-2}]
\ee
and the corresponding evolution eigenvalues $\bar \lambda$ in the topological framing are equal to
\be
\bar \lambda_m = \left(-q^{m-1}A\right)^m
\label{mudef}
\ee

\bigskip

$\bullet \ {\bf Dimensions.}$ The quantum dimensions $\Delta_m$ of representations arising in (\ref{parrep}) are equal to
\be
\label{pDelta}
\Delta_m = \chi_{[r+m,r-m]}={[2m+1]\over [r+m+1]![r-m]!}\prod_{i=0}^{2r-1}D_j\prod_{j=0}^{r-m-1}{D_{j-1}\over D_{r+m+j}}\ ,
\ee
while the quantum dimensions $\bar \Delta_m$ of representations arising in (\ref{aprep}) are
\be
\bar \Delta_m = D_{2m-1}\cdot \left(\prod_{j=0}^{m-2}\frac{D_j}{[j+2]} \right)^{\!\!2}\!\cdot D_{-1}
\label{aDelta}
\ee

\bigskip

$\bullet \ {\bf Universal\ matrix.}$ The third constituent is a universal matrix ${\cal A}$ (we typically use the notation $a_{ij}$ for its matrix elements) which ultimately turns out to be related with the matrix of the quantum Racah coefficients (or, up to a factor, of the $6j$-symbols). In fact, in order to describe all the pretzel links and knots we will need three different universal matrices ${\cal A}$.
After some tedious calculations we have found the following explicit formulas for ${\cal A}$:
\be
\label{RacahN}
a_{km} = \alpha_{km} \, \cdot \, {\cal G} \\
\bar a_{km} = \alpha_{km} \, \cdot \, \bar {\cal G} \label{RacahaN}\\
{\bar{\bar a}}_{km}={\bar\Delta_m\over\Delta_k}a_{mk}
\label{RacahaoN}
\ee
where $\alpha_{km}$ are the coefficients in the $SU_q(2)$ case (i.e. $A=q^2$), which does not differ between the parallel and antiparallel orientations:
\be
\alpha_{km} = (-1)^{r+k+m} [2m+1]\cdot {\Big([k]![m]!\Big)^2\,[r-k]!\,[r-m] !\over [r+k+1]!\,[r+m+1]!}
 \times \nn \\ \times
\ \sum_{j=\text{max}(r+m,r+k)}^{\text{min}(r+k+m,2r)} {(-1)^j\,[j+1]!\over [2r-j]!\ \Big([j-r-k]!\,[j-r-m]!\,[r+k+m-j]!\Big)^2}
\label{Racah2}
\ee
and we introduce the following special functions
\be
{\cal G} &=& \frac{G(r-m)\,G(j+1)}{G(r+k+1)\,G(j-r-m)} \\
\bar {\cal G} &=& \frac{D_{2m-1}}{[2m+1]}\cdot \frac{G(m)^2\,G(j+1)}{G(r+k+1)\,G(r+m+1)\,G(r+k+m-j)} \\ \nn \\
G(n) &=& \dfrac{1}{[n]!}\prod_{i=-1}^{n-2}D_{i}={(A/q;q)_n\over (q;q)_n}
\label{defG}
\ee
where we used the symmetric q-Pochhammer symbol $(A;q)_n=\prod_{j=0}^{n-1}\{Aq^j\}$. At $A=q^N$, $G(n)$ becomes the q-binomial
$\left(\begin{array}{c} N+n-2\\n
\end{array}\right)_q$

Let us note that ${\cal G}$ and $\bar {\cal G}$ are equal to $1$ when $A=q^2$, thus reducing (\ref{RacahN}) and (\ref{RacahaN}) to (\ref{Racah2}). Particular examples of these matrices are given in Appendix B.

Matrix (\ref{RacahN}) satisfies the weighted orthogonality relation
\be\label{orthr1}
\sum_{k=0}^r \ \bar \Delta_k\cdot a_{km}a_{km'} \
= \  \Delta_m\,\delta_{m,m'}.
\ee
The dual relation is
\be
\sum_{m=0}^r\ \frac{a_{km}a_{k'm}}{\Delta_m}
\ =\  \frac{ \delta_{k,k'}}{\bar \Delta_k}.
\label{orthr}
\ee
Matrix (\ref{RacahaN}) also satisfies the orthogonality conditions:
\be\label{orthr2}
\sum_{k=0}^r \bar \Delta_k \cdot \bar a_{km}\bar a_{km'} &=& \bar \Delta_m\cdot \delta_{m,m'}, \nn\\
\sum_{m=0}^r \frac{\bar a_{km}\bar a_{k'm}}{\bar \Delta_m} &=& \frac{ \delta_{k,k'} }{\bar \Delta_k}
\ee
and matrix (\ref{RacahaoN}) satisfies the orthogonality conditions:
\be\label{orthr3}
\sum_{k=0}^r  \Delta_k\cdot \overline{\overline{a}}_{km} \overline{\overline{a}}_{km'}
=\bar\Delta_m\, \delta_{m,m'}\nn
\\
\sum_{m=0}^r  \cfrac{\overline{\overline{a}}_{km} \overline{\overline{a}}_{k'm}}{\bar\Delta_m}
= \frac{\delta_{k,k'}}{\Delta_k}
\ee

The 0th rows of matrices (\ref{RacahN}) and (\ref{RacahaN}) are equal to the quantum dimensions of the corresponding representations (\ref{parrep}) and (\ref{aprep}):
\be\label{o1}
a_{0m} = \Delta_m \\
\bar a_{0m} = \bar \Delta_m
\label{o2}
\ee

Now we specify formula (\ref{poldeco1}) for three possible cases of pretzel knots/links.

\bigskip

\paragraph{Antiparallel odd case.}

Let us consider the case when all parameters $n_1,\dots,n_g$ are odd and all strand into constituent braids are antiparallel. This case is stand-alone and does not mix with any others, i.e. it is impossible to represent knot or link with $n_1,\dots,n_i$ odd antiparallel 2-strand braids and $n_i,\dots,n_g$ odd parallel or even (anti)parallel 2-strand braids. Since for all qualities standing for the antiparallel case we use "bar", we denote parameters in this case as $\overline {n_1},\dots,\overline {n_g}$. Concerning topological classification of this case we can point out the following: if the genus $g$ is odd then the result is a 2-component link, if the genus $g$ is even, the result is a knot. Now let us specify (\ref{poldeco1}) for this particular case:
\be
\dim_qX &=& \Delta_k\nn\\
A_{XY} &=&  \bar{\bar a}_{km} \\
\lambda_Y^{n_i} &=& \bar \lambda_m^{\bar n_i}\nn
\ee
so that formula (\ref{poldeco1}) takes the form
\be
\boxed{ \ \
H^{\overline {n_1},\ldots,\overline  {n_{g+1}}}_R \ \ \ \  \ = \  \sum_{k=0}^r \Delta_k \, \prod_{i=1}^{g+1} \, \sum_{m=0}^r \bar{\bar a}_{km} \bar \lambda_m^{\bar n_i}
}
\label{homfa}
\ee

\paragraph{Other cases.}

All other possible configurations of the pretzel links can be unified into one family with $n_1,\ldots,n_{2g_{||}}$ arbitrary integers associated with the parallel braids and $\overline{n_{2g_{||}+1}},\ldots,\overline{n_{g+1}}$ {\bf even} integers associated with the antiparallel braids. Then, the constituents of (\ref{poldeco1}) are:
\be
&&\dim_qX = \bar \Delta_k \nn\\
&&\left.
\begin{array}{l}
A_{XY} =  a_{km}  \\
\lambda_Y^{n_i} = \lambda_m^{n_i}
\end{array}
\right\} \text{for $n_1,\ldots,n_{2g_{||}}$} \\
&&\left.
\begin{array}{l}
A_{XY} =  \bar a_{km}  \\
\lambda_Y^{n_i} = \bar \lambda_m^{\bar n_i}
\end{array}
\right\} \text{for $\overline{n_{2g_{||}+1}},\ldots,\overline{n_{g+1}}$},\nn
\ee
so that the answer takes the form:
\be
\boxed{ \ \
H^{n_1,\ldots,n_{2g_{||}},
\overline{n_{2g_{||}+1}},\ldots,\overline{n_{g+1}}}_{[r]}
 \ = \ \sum_{k=0}^r \, \bar \Delta_k \cdot \left\{ \ \prod_{i=1}^{2g_{||}} \ \left( \sum_{m=0}^r a_{km} \, \lambda_{m}^{n_i} \right) \, \cdot \, \prod_{j=2g_{||}+1}^{g+1} \left(\sum_{m=0}^r \bar a_{km} \, \bar \lambda_{m}^{\bar n_j}\right) \ \right\} \ \
}
\label{poldeco}
\ee

Thus, our formulas (\ref{homfa}) and (\ref{poldeco}) provide {\bf the explicit answer for arbitrary pretzel link in arbitrary symmetric representation.} These formulas (\ref{poldeco1}) are perfectly consistent with (and, in fact, partly inspired
by) the arbitrary genus results of \cite{GalS} for the Jones polynomials.

The HOMFLY polynomials in the totally antisymmetric representations are obtained by the usual
transposition rule \cite{DMMSS,GS}:
\be
H_{[1^r]}(A,q) = H_{[r]}(A,q^{-1})
\ee

\section{Comments on the main result (\ref{poldeco1})\label{ex}}

\paragraph{Pretzel family.}

The pretzel links and knots provide us with an ample set of examples of the HOMFLY polynomials in all (anti)symmetric representations.
The only examples available so far were: the Whitehead and Borromean rings links \cite{AENV}, the two-strand torus \cite{FGS1,evo} and twist \cite{FGS2,Rama} knots parameterized by one integer each and the double braid unifying these two families and parameterized by two integer numbers \cite{evo}. These families are a tiny part of the whole pretzel family (see s.\ref{PF}). One of the essential points is that the pretzel family includes both thin and think \cite{DGR} knots, while the two-strand torus and twist knots are all thin. The simplest example of the thick pretzel knot is $10_{139} = (4,-1,3,3)$ (see \cite[eq.(49)]{DGR}) in accordance with the Rolfsen tables \cite{katlas}.
This knot can be also obtained from knot $5_2$ by involving a triple braid (see \cite{DMMSS} for details). In s.\ref{PF} we list more patterns from the pretzel family.

\paragraph{Torus in the t-channel = ($\overline{1},\overline{1},\overline{1},\overline{1},\overline{1},\ldots)$.}

The key to understanding the structure of eq.(\ref{homfa}) is to note that
in the particular case, when all $n_i=1$, we actually obtain "in the $t$-channel"
the ordinary two-strand torus link/knot:
${\rm Pretzel}(\bar 1,\ldots, \bar 1) = {\rm Torus}[2,g+1]$, i.e.
\be
H_{[r]}^{(\overline{1},\ldots,\overline{1})} = c_r^{g+1} H_{[r]}^{[2,g+1]}
= c_r^{g+1} \sum_{k=0}^r \Delta_k\lambda_k^{-g-1}
\ee
where $c_r$ is a framing factor, taking into account the difference between
vertical and topological framings.
The "$s$-channel" decomposition formula in this case is
\be
\boxed{
H_{[r]}^{(\overline{1},\ldots,\overline{1})}
= \sum_{k=0}^r \Delta_k\prod_{j=1}^{g+1} \left(\sum_{m=0}^r
\overline{\overline{a}}_{km} \mu_m\right)
}
\label{apaodd}
\ee
which implies
\be
\sum_{m=0}^r \overline{\overline{a}}_{km} \mu_m = c_r \lambda_k^{-1}
\label{amul}
\ee
This is, indeed, the case:
\be
\left.\frac{1+D_1\mu_1}{D_0}\right|_{\mu = -A} = -Aq= c_1, \ \ \ \
\left.\frac{1-D_{-1}\mu_1}{D_0}\right|_{\mu = -A} = \frac{1}{Aq} = c_1\cdot(-q^{-2})
= c_1\lambda_1^{-1}
\ee
In fact, along with the orthogonality conditions (\ref{RacahaoN}), this requirement allows one to restore the whole matrix ${\cal A}$ in this case.

\bigskip

{\sl
In the next paragraph we consider time-dependent quantities, thus the label $*$,
referring to restriction to topological locus, which was omitted throughout the main text,
is restored.
}

\paragraph{Generalizing the Rosso-Jones formula.}

Let us return to the Rosso-Jones formula (\ref{RJ}).
In the case of symmetric representations $R=[r]$, it can be written in the form
 \be
 {\cal H}^{[m,n]}_{[r]}\{p\} = q^{\frac{2n}{m}\hat W_{[2]}}\,\hat\pi \ \chi_{[r]}\{p\}^m
\label{RJtor}
 \ee
with the operator $\hat \pi$ changing sign of the odd character $\chi_m$
\be
\hat\pi \ \chi_{m}\{p\} = (-)^m\chi_{m}\{p\}
\ee
where $m$ labels representations $Q_m$ arising in the two decompositions
\be
[r]\otimes[r] = \oplus_{m=0}^r [r+m,r-m]
\ee
and
\be
[r]\otimes\overline{[r]} = \oplus_{m=0}^r [2m,m^{N-2}]
\ee

Clearly, our (\ref{poldeco}) implies an extension of (\ref{RJtor})
to arbitrary pretzel links/knots:
$$
H^{n_1,\ldots,n_{2g_{||}},
\overline{n_{2g_{||}+1}},\ldots,\overline{n_{g+1}}}_{[r]}\{p\} \ =
\ \ \ \ \ \ \ \ \ \ \ \ \ \ \ \ \ \ \ \  \ \ \ \ \ \ \ \ \ \ \ \ \ \ \ \ \ \ \ \
$$
\vspace{-0.6cm}
\be
= \ \left(\otimes_{I=1}^{g+1}q^{n_I\hat W_{[2]}(p^{(I)})}\right)
 \sum_{k=0}^r \, \bar \Delta_k \cdot \left\{ \ \prod_{i=1}^{2g_{||}} \ \left( \sum_{m=0}^r {a_{km}\over\chi_m^\ast}\ \chi_m\Big(p^{(i)}\Big) \,\right) \, \cdot \, \prod_{j=2g_{||}+1}^{g+1} \left(\sum_{m=0}^r {\bar a_{km}\over\bar\chi_m^\ast} \  \bar\chi_m\Big(p^{(j+2g_{||})}\Big)\right) \ \right\} \ \
\ee
A similar extension exists for (\ref{homfa}).

Significant difference from (\ref{RJtor}) is that the rotation matrices $a_{km}$
and $\bar a_{km}$ depend on the representation $[r]$, and it is a challenging problem
to encode this dependence into the action of some operator.

Also note that beyond the topological locus
\be
{\cal H}^{(n_1,n_2)}\{p^{(1)},p^{(2)}\} \  \neq \ {\cal H}^{(n_1+n_2)}\{p\}
\ee
-- the two sides even depend on different sets of time-variables.

\paragraph{Extension to superpolynomials and to non-symmetric representations.}
As known since \cite{IMMMfe}, generalization of
formulas like (\ref{homfa}) and (\ref{poldeco}) to (anti)symmetric  {\it super}polynomials
is straightforward. However, constructing the superpolynomials and another problem that can be solved by an immediate extension of these formulas, that is, constructing the HOMFLY polynomials
in other representations will be considered elsewhere. Presently the best, what is known beyond arbitrary torus knots
(where Rosso-Jones formula \cite{RJ,Che} provides generic answer in arbitrary
representation) are twist knots in representation $[21]$, see \cite{Ano21},
\cite{germ21} and, finally, \cite{MMM21}, see also \cite{DanChe} for
a family of torus descendants.
It is (\ref{poldeco1}) that allowed us to make a far-going conjecture \cite{GMMMS}
about a generalization of Rosso-Jones formula to all representations of
genus-$g$ knots; it, however, remains to be checked.
An even more challenging question is about associated generalization of the
eigenvalue matrix model (\ref{TBEM}), currently it is available only for twist knots \cite{Almamo}.

\paragraph{$A$-polynomials.}
One can study the dependence of the constructed HOMFLY polynomials (\ref{homfa}) and (\ref{poldeco}) on spin of the representation
(or representations in the case of links). One of the ways to describe this dependence is to derive difference equations
with respect to the spin variables. There are various types of these relations \cite{MMeqs}, some of them are very easy to observe,
other ones are usually much more complicated but instead they can be related to the volume conjecture \cite{vc} and their
``quasiclassical" limit is given by the $A$-polynomial \cite{Gar} (the so-called AJ-conjecture) and, for this reason, the equations are called "quantum $A$-polynomials". They can be found with the help of computer programs
implementing Zeilberger's algorithm for the
hypergeometric sums \cite{Zeil,Koorw}. Since the HOMFLY polynomials in any symmetric representations were known so far only for a few cases (see above), only in those cases the quantum $A$-polynomial was calculated. Our results (\ref{homfa}) and (\ref{poldeco}) open a road for obtaining many more $A$-polynomials, though these expressions literally are not suitable and still have to be reshuffled:
they have no form of a $q$-hypergeometric polynomial and the existing software
implementing Zeilberger's algorithm for the
hypergeometric sums \cite{Zeil,Koorw}
can not be immediately used.

\section{Matrices $a_{km}$ and $\bar a_{km}$ as universal Racah matrix
\label{ARacah}}

In the case of Jones polynomials (i.e., for $A=q^2$),
the simplest matrix ${\cal A}$ turns into
\be
{\cal A}_1 = \left(\begin{array}{cc}  1 & [3] \\ 1 & -1 \end{array}\right)
\ee
and an immediate desire is to compare it with the celebrated fusion (mixing) matrix
\be
S=\frac{1}{[2]}\left(\begin{array}{cc}  1 & \sqrt{[3]} \\ \sqrt{[3]} & -1 \end{array}\right)
\ee
which recently appeared in many places,
from modular transformation of the simplest Virasoro conformal block in \cite{Gal}
to elementary three-strand knot calculations in \cite{MMMknots}.
This similarity turned out to be not a simple coincidence, but a manifestation
of general fact: in full generality the matrices ${\cal A}$ in our formulas
for the genus-$g$ knot polynomials are nothing but a simple rescaling of
the Racah matrices from representation theory of quantum groups,
of which $S$ is just the simplest example.
This fact, what came for us as a result of tedious calculations,
was announced in a separate paper \cite{GMMMS}.
Though this is nearly obvious after being discovered
and is spectacularly confirmed by the derivation of eqs.(\ref{RacahN})-(\ref{RacahaoN}),
in this section we provide a little more details and comments.

First of all, in variance with the Jones case, where the relevant group is $SU_q(2)$
and the Racah matrices are long known from \cite{Racahsl2},
in the HOMFLY case one needs generic $SU_q(N)$ matrices ${\cal A}$
which depend on $A$. Therefore, they are {\it universal} objects, interpolating
between the Racah matrices for particular $SU_q(N)$ at $A=q^N$.
Not much was known about such quantities until recently, fortunately,
the very recent \cite{indsRacah} provides the needed information.
Second, in the HOMFLY case the set of allowed representations is wider than that
in the Jones case: even if one restricts considerations to symmetric representations,
their conjugates unavoidably enter the game, and they are no longer the same,
as they were for $SU_q(2)$.

Having this said, let us return to our main formula (\ref{poldeco1}) which was conjectured in \cite{GMMMS} yet
\be\label{rep}
H^{n_1,\ldots,n_{g+1}}_R \ \ \ \  \ = \  \sum_{\X} \dim_q\X \, \prod_{i=1}^{g+1} \, \sum_\Y {\cal A}_{\X\Y}\lambda_\Y^{n_i}
\ee
and comment on it in a little more detail. In fact, this formula naturally generalizes
to $l$ different representations in the case of $l$-component link \cite{GMMMS} (see Figure 6).

\begin{itemize}

\item[{\bf 1.}] In this formula, one can understand under the calligraphic index either $X$ or $\bar X$ and similarly for $\Y$
so that $\bar\Delta_X\equiv\Delta_{\bar X}$,
$\bar\lambda_X\equiv\lambda_{\bar X}$ etc. Then, there are three possibilities: when in (\ref{rep}) enter $X$ and $\bar Y$,
$\bar X$ and $\bar Y$, $\bar X$ and $Y$. Accordingly, there are three different matrices ${\cal A}_{X\bar Y}$, ${\cal A}_{\bar X\bar Y}$
and ${\cal A}_{\bar X Y}$ which correspond to (\ref{RacahN}), (\ref{RacahaN}) and (\ref{RacahaoN}).

\item[{\bf 2.}] The three orthogonality conditions (\ref{orthr1})-(\ref{orthr3}) satisfied by the matrices ${\cal A}$
can be rewritten in these terms as the single equation
\be
\sum_{\X}  {\rm dim}_q\X\cdot
{\cal A}_{\X\Y}\cdot
{\cal A}_{\X\Y'}
={\rm dim}_q\Y \,\delta_{\Y,\Y'}
\ee
and similarly for the dual one (\ref{orthr}).
This means that the relation to the orthonormal Racah matrix $S$ is
\be
{\cal A}_{\X\Y} =  \sqrt{\frac{{\rm dim}_q\Y}{{\rm dim}_q\X}}\, S_{\X\Y}
\ee
After such a rescaling our formulas (\ref{RacahN}) and (\ref{RacahaN})
seem to be in a perfect agreement with the conjectures of \cite{indsRacah},
thus justifying/supporting our suggested identification of ${\cal A}$
as the rescaled Racah matrices.

\item[{\bf 3.}] Note that there are exactly three possible Racah matrices when all the representations are either $R$ or $\bar R$:
$\ S\!\left(\begin{array}{cc} R & \overline{R} \\ R & \overline{R} \end{array}\right)$,
$S\!\left(\begin{array}{cc} \overline{R} & \overline{R} \\ R & R \end{array}\right)$ and
$S\!\left(\begin{array}{cc} \overline{R} & R \\ R&\overline{R} \end{array}\right)$.

These three cases correspond to the three types of the matrices ${\cal A}$ discussed above: ${\cal A}_{X\bar Y}$, ${\cal A}_{\bar X Y}$
and ${\cal A}_{\bar X \bar Y}$. The first two are transposed to each other (by the general properties of the Racah matrices) and the third one is symmetric.
This also explains why the factor $\bar {\cal G}$ in (\ref{defG}) is symmetric in $k$ and $m$, while ${\cal G}$ is not:
in the latter case $k$ and $m$ label different representations, $k \in [r]\otimes \overline{[r]}$,
$m\in [r]\otimes[r]$, while in the former case both $k,m\in[r]\otimes\overline{[r]}$.

\item[{\bf 4.}] One can make use of the additional fact that the first line of the matrices
${\cal A}_{\X\Y}$, associated with the singlet representation $X=\overline{\emptyset}$,
consists just of the quantum dimensions $\chi_\Y={\rm dim}_q\Y$, (\ref{o1}), (\ref{o2})
and rewrite (\ref{rep}) through the orthonormal Racah matrix $S$ in another form:
\be
\boxed{
H^{n_1,\ldots,n_{g+1}}_R=
\sum_\X  \left({\rm dim}_q\X\right)^{\frac{1-g}{2}}  \prod_{i=1}^{g+1}
\left( \sum_\Y  S_{\X\Y}S_{\overline{\emptyset} \Y} \lambda_\Y^{n_i}\right)
}
\label{genRHans2}
\ee
In result, the contributions of parallel, even antiparallel and
odd antiparallel braids are respectively
$S_{\bar XY}S_{\overline{\emptyset} Y}$, $S_{\bar X\bar Y}S_{\overline{\emptyset} \bar Y}$ and $S_{X\bar Y}S_{\overline{\emptyset} \bar Y}$.
In the case of $g=1$ (torus knots/links) the factor ${\rm dim}_q\X$
is absent and one can sum over $\X$, using the orthonormality condition
$\sum_\X S_{\X\Y}S_{\X\Y'}=\delta_{\Y\Y'}$, to get just the Rosso-Jones formula
in the form of \cite{GalS}
\be
H^{(n_1,n_2)} = \sum_\Y S_{\emptyset \Y}^2 \lambda_\Y^{n_1+n_2}
\ee
The structure of (\ref{genRHans2}), involving a sum
with a weight, which is the power $2-2g$ (the Euler characteristics of the genus $g$ Riemann surface) of representation dependent quantity resembles the Frobenius formula \cite{Dijk}, typical for
topological (cohomological) models.

\item[{\bf 5.}]
At a deeper level, the relation of (\ref{genRHans2}) to the conformal block calculus
of \cite{inds} and \cite{GalS} remains a mystery. It can be schematically realized with a toric conformal block picture, see Figure 6.
The occurrence of the {\it toric}
blocks can seem natural for the pretzel family, but exact appearance,
and the very possibility to derive (\ref{genRHans2}) from consideration
of the {\it spherical} blocks, as done in \cite{GalS} implies
some interrelation in the style of Verlinde formulas,
which needs to be put in a more precise form.

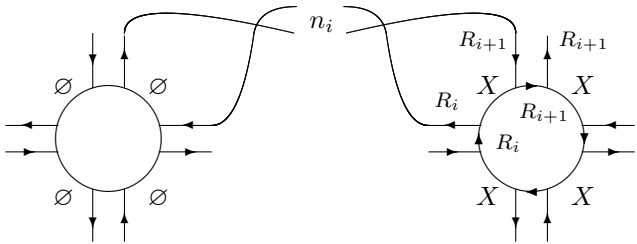
\begin{figure}[h!]\label{tcf}
\begin{picture}(100,115)(-300,-50)
\put(0,0){\circle{40}}
\put(-19,-5){\line(-1,0){20}}  \put(-30,-5){\vector(1,0){2}}
\put(-19,5){\line(-1,0){20}}   \put(-31,5){\vector(-1,0){2}}
\put(19,-5){\line(1,0){20}}     \put(30,5){\vector(-1,0){2}}
\put(19,5){\line(1,0){20}}      \put(31,-5){\vector(1,0){2}}
\put(-6,19){\line(0,1){21}}      \put(-6,30){\vector(0,-1){2}}
\put(6,19){\line(0,1){20}}      \put(6,32){\vector(0,1){2}}
\put(-6,-19){\line(0,-1){20}}    \put(-6,-32){\vector(0,-1){2}}
\put(6,-19){\line(0,-1){20}}     \put(6,-30){\vector(0,1){2}}
\put(0,20){\vector(1,0){2}}
\put(20,0){\vector(0,-1){2}}
\put(0,-20){\vector(-1,0){2}}
\put(-20,0){\vector(0,1){2}}
\put(-40,12){\mbox{  {\footnotesize $R_i$}  }}
\put(-17,-4){\mbox{  {\footnotesize $R_i$}  }}
\put(-31,35){\mbox{  {\footnotesize $R_{i+1}$}  }}
\put(7,35){\mbox{   {\footnotesize $R_{i+1}$}  }}
\put(-8,8){\mbox{    {\footnotesize $R_{i+1}$}  }}
\put(15,17){\mbox{$X$}}
\put(15,-25){\mbox{$X$}}
\put(-21,-25){\mbox{$X$}}
\put(-21,17){\mbox{$X$}}
\qbezier(-6,40)(-6,55)(-70,40)
\qbezier(-39,5)(-50,5)(-55,40)
\qbezier(-55,40)(-57,50)(-71,50)
\qbezier(-154,40)(-154,55)(-90,40)
\qbezier(-121,5)(-110,5)(-105,40)
\qbezier(-105,40)(-103,50)(-89,50)
\put(-84,42){\mbox{$n_i$}}
\put(-160,0){\circle{40}}
\put(-179,-5){\line(-1,0){20}}  \put(-190,-5){\vector(1,0){2}}
\put(-179,5){\line(-1,0){20}}   \put(-191,5){\vector(-1,0){2}}
\put(-141,-5){\line(1,0){20}}     \put(-130,5){\vector(-1,0){2}}
\put(-141,5){\line(1,0){20}}      \put(-129,-5){\vector(1,0){2}}
\put(-166,19){\line(0,1){21}}      \put(-166,30){\vector(0,-1){2}}
\put(-154,19){\line(0,1){20}}      \put(-154,32){\vector(0,1){2}}
\put(-166,-19){\line(0,-1){20}}    \put(-166,-32){\vector(0,-1){2}}
\put(-154,-19){\line(0,-1){20}}     \put(-154,-30){\vector(0,1){2}}
\put(0,20){\vector(1,0){2}}
\put(20,0){\vector(0,-1){2}}
\put(0,-20){\vector(-1,0){2}}
\put(-20,0){\vector(0,1){2}}
\put(-145,17){\mbox{$\varnothing$}}
\put(-145,-25){\mbox{$\varnothing$}}
\put(-181,-25){\mbox{$\varnothing$}}
\put(-181,17){\mbox{$\varnothing$}}
\end{picture}
\caption{Conformal block representation of formula (\ref{genRHans2}). Here the generic case of different representations $R_i$
(in the case of link) is drawn, it corresponds to
$\sum_{\bar X\, \in\, \bigcap_i R_i\otimes \bar R_i}  {\rm dim}_q\bar X
\cdot \prod_{i=0}^{g} \left(
\sum_{\bar Y_i\in R_i\otimes \bar R_{i+1}}
{\cal A}_{\bar X\bar Y_i}\Big(^{_{\bar R_i \  R_{i+1}}}_{^{R_i \ \bar R_{i+1}}}\Big)
\cdot \lambda_{\bar Y_i}^{n_i}\right)
$,
or other two matrices ${\cal A}$ depending on the direction of arrows in the picture}
\end{figure}

\item[{\bf 6.}] For generic representations one has:
$R_1\otimes R_2 = \oplus\  X\otimes V_X$.
When the multiplicities of all $X$ are unities, as in the case
of a product of two symmetric representations and/or of their conjugates,
${\rm dim}_q\,X$ is just a number, dimension of the representation $X$.
In this case, (\ref{genRHans2}) is symmetric under arbitrary permutations of $n_i$
(if all $R_i$ are the same), i.e. there is the enhanced symmetry.
However, when multiplicity of $X$ is non-trivial, i.e. $V_X$ is a vector space
of non-unit  dimension, then the matrix ${\cal A}_{XY}$ is a vector in $V_X$
and ${\rm dim}_q\, X$ in (\ref{genRHans2}) is rather a multi-linear operation
$V_X^{\otimes (g+1)} \longrightarrow 1$, which does not need to be totally
symmetric: only the cyclic symmetry is needed.
This is in accordance with the fact that a non-cyclic permutation of $n_i$
converts a knot/link into a {\it mutant}, undistinguishable
by symmetrically-colored knot
polynomials, but separable by those in non-(anti)symmetric representations.
\end{itemize}

\section{Which knots and links belong to the genus-$g$ family\label{PF}}

So far there were just two {\it families} with explicitly known (anti)symmetric HOMFLY polynomials:
torus knots \cite{RJ,DMMSS} and double braids \cite{evo}. It is this second family that we enormously extend in the present paper.
The double braids are $ \ \text{Pretzel}:(-1,\overline{2k},n)=(\overline{n},\underbrace{\overline{1},\ldots,\overline{1}}_{2k})$ and contain twist knots $(n=-1)$, in particular the figure-eight knot $4_1$, which was the first example beyond torus knots, studied  in \cite{IMMMfe}.
Our formulas are of course consistent with the results of \cite{RJ,DMMSS} and \cite{evo}, but \cite{evo} and \cite{artdiff}
contains more: differential expansions, which still needs to be studied in generic pretzel case.

Below we list some tests which we did to examine our main conjecture (\ref{poldeco1}). Also we tabulate some knots and links which belong to the family of 2-strand genus-g knots.

\paragraph{1.} For some particular examples we have checked that our formula (\ref{poldeco1}) reproduces the HOMFLY polynomials in symmetric representations for:

\begin{itemize}
\item
twist knots \cite{IMMMfe,FGS2,Rama,evo}: $T^{(k)}  = (\bar 1,\bar 1,\overline{2k-1}) = (-1,-1,\overline{2k})$ (we remind that $T^{(-k)} = (2k+2)_1$   for $k>0$, $T^{0}=$ unknot, $T^{(1)} = 3_1 $,  $T^{(k)} = (2k+1)_2 $ for $k>1$);
\item
torus knots and links \cite{FGS1,evo}:  $[2,n] = (n,0)=(1,n-1)$, $[3,4] = 8_{19} =  (3,3,\overline{-2})$, $[3,5]=(5,3,\overline{-2})$, all other are {\it not} pretzel knots;
\item
double braid \cite{evo}: $(-1,\overline{2k},n)=(\overline{n},\underbrace{\overline{1},\ldots,\overline{1}}_{2k})$.
\item
Whitehead link \cite{AENV} (in this case two {\it different} representations are allowed on the two different components of the link): $L_5a_1=(2,\overline{2},1)$.
\end{itemize}

\paragraph{2.} Special polynomials $H_R^{\mathcal{K}}(A,q=1)$ provided a very good test due to the factorization property \cite{DMMSS,Zhu}:
\be
\sigma_R(A) = \lim_{q\longrightarrow 1}\ \frac{H_R(q.A)}{\chi_R(q,A)} = \Big(\sigma_{[1]}(A)\Big)^{|R|}
\label{factorspe}
\ee

\paragraph{3.} Alexander polynomials $H_R^{\mathcal{K}}(A=1,q)$ also have a simple representation dependence for hook diagrams which include symmetric representations \cite{IMMMfe}
\be
\frac{H_{R}}{\chi_R}(A=1,q) = \frac{H_{[1]}}{\chi_{[1]}}(A=1,q^{|R|})
\label{alexrel}
\ee
and also provided a very good test for our results.

\paragraph{4.} Another powerful tool to test the HOMFLY polynomials in any representation, which we used here, is expansion through the Vassiliev invariants and trivalent diagrams. We shall not explain this expansion here and refer our readers to the literature \cite{Vass}.

\paragraph{5. Table: Rolfsen and Thistlethwaite  vs.  pretzel \cite{katlas}.}

Understanding if the given knot/link belongs to the pretzel family
is not quite a trivial exercise.
However, since we now possess generic expression for pretzel symmetric HOMFLY,
this can be done systematically,
by comparing our results with the polynomials in \cite{katlas}
(note that there is just one coincidence between fundamental HOMFLY for
up-to-ten-intersection knots --
for $5_1$ and $10_{132}$).
Result of this analysis is the following list.
We use condensed notation $1^5=1,1,1,1,1$ and
also do not distinguish between knots and their mirrors, when the signs
in all crossings are reversed. Symbol "$\leftrightsquigarrow$" stands for knot mutation.

\renewcommand{\arraystretch}{1.8}
\be
\begin{array}{|c|c||c|c||c|c||c|c||c|c|}
\hline
\multicolumn{10}{|c|}{\textbf{Knots, \texttt{\#}crossings\,=\,3..8}}\\
\hline
{\bf 3_1} & (3,0) & {\bf 7_1} & (7, 0) & {\bf 8_1} & (1,\bar 6,1) & {\bf 8_8} & (2, -3, 1, 1, 1, 1)  & {\bf 8_{15}} & (2,3,3, -1, -1, -1)    \\
\hhline{|-|-|-|-|-|-|-|-|-|-|}
{\bf 4_1} & (1,\bar 2, 1)  & {\bf 7_2} & (\bar 5, \bar 1,\bar 1) & {\bf 8_2} & (5,\bar 2,1)  &  {\bf 8_9} & (4, -3, -1, -1)  & {\bf 8_{16}} &     \\
\hhline{|-|-|-|-|-|-|-|-|-|-|}
{\bf 5_1} & (5,0)  & {\bf 7_3} & (4,1,1,1) & {\bf 8_3} & (1,1,\bar 4,1,1)  &   {\bf 8_{10}} & (2, -3, 1, 3)  &  {\bf 8_{17}} &     \\
\hline
{\bf 5_2} & (\bar 3,\bar 1,\bar 1)  &  {\bf 7_4} & (\bar 3,\bar 3,\bar 1) & {\bf 8_4} & (3,\bar 4,1)  & {\bf 8_{11}} & (-\bar 3, \bar 1, \bar 1, \bar 3, \bar 1)  & {\bf 8_{18}} &       \\
\hhline{|-|-|-|-|-|-|-|-|-|-|}
{\bf 6_1} & (\bar 5,-\bar 1,-\bar 1)  &  {\bf 7_5} & (3,2,1,1) & {\bf 8_5} & (3,\bar 2,3)  & {\bf 8_{12}} &   &   {\bf 8_{19}} & (3,-\bar 2,3)   \\
\hline
{\bf 6_2} & (3,\bar 2, 1)  &  {\bf 7_6} & (-3,1,\bar 2, 1,1) & {\bf 8_6} & (1,3,\bar 2,1,1)  &  {\bf 8_{13}} & (-\bar 4,-3, 1, 1, 1)  &  {\bf 8_{20}} & (3,\bar 2,-3) \\
\hline
{\bf 6_3} & (2,-3,1,1)  &  {\bf 7_7} & (-\bar 3, \bar 1, -\bar 3, \bar 1, \bar 1) & {\bf 8_7} & (4, -3, 1, 1)  & {\bf 8_{14}} &   &  {\bf 8_{21}} & (2, -3, 1, -3)     \\
\hline
\multicolumn{10}{|c|}{\textbf{Links}}\\
\hline
{\bf L_2a_1} & (2,0) & {\bf L_6a_1} & (2,1,1,2) & {\bf L_6a_4} & ? &  &   &  &     \\
\hline
{\bf L_4a_1} & (4,0) & {\bf L_6a_2} & (3,1,1,1) & {\bf L_6a_5} & (2,\bar 2, 2) &  &   &  &     \\
\hline
{\bf L_5a_1} & (2,\bar 2, 1) & {\bf L_6a_3} & (6,0) & {\bf L_6n_1} & (-2,-\bar 2,2) &  &   &  &     \\
\hline
\end{array}
\nn
\ee

\be
\begin{array}{|c|c||c|c||c|c||c|c||c|c|}
\hline
\multicolumn{10}{|c|}{\textbf{Knots, \texttt{\#}crossings\,=\,9}}\\
\hline
{\bf 9_1} & (9,0) & {\bf  9_{11}} & (-5,-2, 1^4) & {\bf 9_{21}} &  & {\bf 9_{31}} &  & {\bf 9_{41}} &     \\
\hhline{|-|-|-|-|-|-|-|-|-|-|}
{\bf 9_2} & (\bar 1,\bar 7,\bar 1)  & {\bf 9_{12}} & (-3, 1, 1, 1, \bar 4) & {\bf 9_{22}} &   &  {\bf 9_{32}} &   & {\bf 9_{42}} &     \\
\hhline{|-|-|-|-|-|-|-|-|-|-|}
{\bf 9_3} & (6, 1, 1, 1)  & {\bf 9_{13}} & (1, 3, 1, 1, -\bar 4) & {\bf 9_{23}} &   &   {\bf 9_{33}} &   &  {\bf 9_{43}} &     \\
\hline
{\bf  9_4} & (4,1^5)  &  {\bf 9_{14}} & (-\bar 5, -\bar 3, \bar 1, \bar 1,\bar 1) & {\bf 9_{24}} & (-2, -3, 3, 1^3)  & {\bf 9_{34}} &   & {\bf 9_{44}} &       \\
\hhline{|-|-|-|-|-|-|-|-|-|-|}
{\bf 9_5} & (-\bar 1, -\bar 3, -\bar 5)  &  {\bf 9_{15}} &  & {\bf 9_{25}} &   & {\bf 9_{35}} &  (\bar 3, \bar 3, \bar 3)  &   {\bf 9_{45}} &    \\
\hline
{\bf  9_6} & (2, 1, 5, 1)  &  {\bf 9_{16}} & (\bar 2, 1, 3, 3) & {\bf 9_{26}} &   &  {\bf 9_{36}} &   &  {\bf 9_{46}} & (\bar 3, -\bar 3, \bar 3) \\
\hline
{\bf  9_7} & (2,3,1^4)  &  {\bf 9_{17}} & (\bar 3, \bar 3, -\bar 1^5) & {\bf 9_{27}} &   & {\bf 9_{37}} & (-\bar 3, -\bar 3, \bar 3, \bar 1, \bar 1)  &  {\bf 9_{47}} &      \\
\hline
{\bf 9_8} & (-2, -3, 1^6)  &  {\bf 9_{18}} &  & {\bf 9_{28}} & (2, -3, -3, 1^3)  & {\bf 9_{38}} &   &  {\bf 9_{48}} & (-\bar 3, -\bar 3, -\bar 3, \bar 1, \bar 1)     \\
\hline
{\bf 9_9} & (-4, 1, -5, 1)  &  {\bf 9_{19}} &  & {\bf 9_{29}} &   & {\bf 9_{39}} &   &  {\bf 9_{49}} &      \\
\hline
{\bf  9_{10}} & (\bar 3, \bar 3, \bar 1, \bar 1, \bar 1)  &  {\bf  9_{20}} & (4, 3, -1^4) & {\bf 9_{30}} &   & {\bf 9_{40}} &   &   &      \\
\hline
\end{array}
\nn
\ee

\centerline{
$
\begin{array}{|c|c||c|c||c|c||c|c|}
\hline
\multicolumn{8}{|c|}{\textbf{Knots, \texttt{\#}crossings\,=\,10}}\\
\hline
{\bf 10_1} & (\bar 1, \bar 7, -\bar 3)  &{\bf 10_{16}} & (\bar 3, \bar 1, -\bar 5, \bar 1, \bar 1)
&{\bf 10_{61}} & (3, 3, \bar 4)
& {\bf 10_{126}} & (\bar 2,-5,3)\bigvee (-2, 3, -5, 1)     \\
\hhline{|-|-|-|-|-|-|-|-|}
{\bf 10_2} & (2, -7, -1, -1)&   {\bf 10_{17}} & (4, -5, 1, 1)
& {\bf 10_{62}} & (4, -3, 1, 3)& {\bf 10_{127}} & (\bar 2,5,3)\bigvee (2, -5, -3, 1) \\
\hhline{|-|-|-|-|-|-|-|-|}
{\bf 10_3} & (\bar 1, \bar 5, -\bar 5)  &  {\bf 10_{19}} & (\bar 4,5,-1,-1,-1)
&{\bf 10_{63}}& (\bar 4,-3,-3,1,1) & {\bf 10_{129}} & (2, 1, 1, -3, 1, 1)  \\
\hhline{|-|-|-|-|-|-|-|-|}
{\bf 10_4} & (-\bar 7,\bar 1,\bar 1,\bar 1,\bar 1)  &  {\bf 10_{20}} & (-2, 1, 3, 1^5)
& {\bf 10_{64}}& (-4, 3, 3, 1)  &  {\bf 10_{139}} &  (4, -1, 3, 3) \\
\hline
{\bf 10_5} & (-2, 7, -1, -1) &  {\bf 10_{21}} & (-\bar 3,\bar 3, \bar 1^5)
&{\bf 10_{65}} &  (\bar 4,3,-3,-1,-1) &{\bf 10_{140}} &  (-3, 3,\bar 4)    \\
\hhline{|-|-|-|-|-|-|-|-|}
{\bf 10_6} & (-\bar 2,-5, -1^3 )&  {\bf 10_{22}} & (-4, 1, 1, 3, 1, 1)
&{\bf 10_{69}} &  (4,3,3,-1,-1,-1) & {\bf 10_{141}} &  (4, -3, -3, 1) \\
\hline
{\bf 10_7} & (-\bar 3,\bar 1,\bar 5,\bar 1,\bar 1)  & {\bf 10_{28}} & (\bar 4,3,-1^5)
& {\bf 10_{74}} & (-\bar 3,\bar 1,\bar 3,\bar 3,\bar 1)& {\bf 10_{142}} &  (3, 3, -\bar 4)     \\
\hline
{\bf 10_8} & (-6, 1^5)   &  {\bf 10_{34}} & (2, -3, 1^6)
& {\bf 10_{76}} &  (1, 3, 3, 1, \bar 2)  &  {\bf 10_{143}} &  (-4, 3, 1, -3)   \\
\hline
{\bf 10_9} & (6, -3, -1, -1) &  {\bf 10_{46}} & (-2, 3, 5, 1)
& {\bf 10_{77}} & (2, -3, 1, 3, 1, 1) &{\bf 10_{144}} &(\bar 4,3,3,-1,-1)    \\
\hline
{\bf 10_{11}} & (3, 1, 1, 1, \bar4)  &  {\bf 10_{47}} & (2, -3, 5, 1)
&{\bf 10_{78}}&  (\bar 2,-3,-3,1,1,1,1)  &&\\
\hline
{\bf 10_{12}} & (4, -3, 1, 1, 1, 1) &  {\bf 10_{48}} & (2, -5, 1, 3)
& {\bf 10_{124}} & (\bar 2,-5,-3) \bigvee (2, -1, 5, 3)  & &  \\
\hline
{\bf 10_{15}} & (-2, -1, 5, -1^3)   & {\bf 10_{49}} &   (\bar 2,-5,-3,1,1)
& {\bf 10_{125}} & (\bar 2,5,-3) \bigvee (2, -5, -1, 3) &&     \\
&&& (2,5,3,-1^3) && &&  \\
\hline
\end{array}
$
}

\bigskip

\be
\begin{array}{|c|c|c|c|c|}
\hline
\multicolumn{5}{|c|}{\textbf{Mutants, \texttt{\#}crossings\,=\,11}}\\
\hline
{\bf 11^a_{44}} & (-3,3,2,1,1,-3) &\leftrightsquigarrow &  {\bf 11^a_{47}} & (3,-3,2,1,1,-3) \\
\hline
{\bf 11^a_{57}} & (\bar 2,1,3,3,-3) &\leftrightsquigarrow  &  {\bf 11^a_{231}} & (\bar 2,1,3,-3,3) \\
\hline
{\bf 11^n_{71}} & (\bar 2,-3,3,-3,1) &\leftrightsquigarrow &  {\bf 11^n_{75}} & (\bar 2,3,-3,-3,1) \\
\hline
{\bf 11^n_{73}} & (2,3,-3,-3)  & \leftrightsquigarrow & {\bf 11^n_{74}} & (2,-3,3,-3) \\
\hline
{\bf 11^n_{76}} & (2,3,3,-3)  &\leftrightsquigarrow &  {\bf 11^n_{78}} & (2,3,-3,3) \\
\hline
\end{array}
\nn
\ee

\bigskip

\section{Conclusion}

In this paper we reported the results about the HOMFLY polynomials
for the pretzel knots, which are a natural
generalization of the torus knots from $g=1$ to arbitrary genus $g$,
for which an exhaustive answer like the Rosso-Jones formula can presumably be found.

Indeed, we found a well-structured exhaustively explicit answer
for arbitrary $g$ in all (anti)symmetric representations,
and indeed the Rosso-Jones formula arises as its very special case.
Not surprisingly, this  general  answer involves more than just
quantum dimensions, but also the Racah matrices, however, in the
absolutely minimal way. As a byproduct of our calculation, an explicit
formula for Racah matrices was found in symmetric representations,
which is completely in accord with the recent result in \cite{indsRacah}.
A stronger conjecture about generic representations,
formulated in \cite{GMMMS} on the base of the present paper,
needs more work to be checked.
In an accompanying paper \cite{GalS} some parallel evidence is
obtained by different method for the Jones polynomials: that work
served as a major inspiration for some of the above calculations.

Further work in this direction seems to be very promising and
can lead to considerable extension of the set of known knot
polynomials.
An absolutely new kind of decomposition of knot polynomials
into the Racah rotated elementary HOMFLY,
as well as emerging a partly expected connection to
the (modular transformations of the) {\it toric} conformal blocks
requires better understanding, perhaps, as a kind of monopole/brane
duality, and suggests various generalizations and implications.
All this can open a new intriguing chapter in the theory of
knot polynomials.

\bigskip

There are five obvious exercises to do, once the
full evolution induced answer is known:
\begin{itemize}
\item[(i)] to derive differential expansions {\it a la} \cite{IMMMfe,artdiff},
\item[(ii)] to find equations w.r.t. the  $r$-parameter \cite{Gar,MMeqs},
\item[(iii)] to study the large-$r$ (Kashaev or volume conjecture) limit \cite{vc},
\item[(iv)] to use it to built a matrix model {\it a la} \cite{Almamo},
\item[(v)] to perform $\beta$-deformation \cite{betadefo} {\it a la} \cite{DMMSS}, \cite{supHL}
and \cite{IMMMfe,FGS1,FGS2,Rama}, i.e. to promote HOMFLY to superpolynomials {\it without}
explicit application of the  sophisticated Khovanov-Rozansky construction
\cite{KhR} and even of its simplified modern substitutes \cite{DM3}.
\end{itemize}
Also straightforward should be generalization to various extensions of the pretzel family,
like combinations of multi-strand braids and their "iterations" {\it a la}
\cite{DanChe}.
These considerations will be reported elsewhere.

\bigskip

A really difficult task is going from (anti)symmetric to generic representations.
If one believes that the conjecture (\ref{rep}) or something similar holds for them,
the problem is actually about the generic Racah matrices, which is a kind of
a classical hard problems in group theory.
Still, as the results in the present paper demonstrate once again,
study of the knot polynomials can provide a new powerful tool to attack such
old  problems: after the answer for the generic symmetric Racah coefficients
appeared very easily in this way, one can anticipate insights about other
representations as well.
For some yet-non-systematic considerations of non-symmetric colored knot polynomials
beyond the torus links see \cite{Ano21,Rama2,germ21,MMM21}.

\section*{Acknowledgements}

We would like to thank A.Anokhina, S.Arthamonov, D.Galakhov, D.Melnikov and And.Morozov for fruitful discussions.

Our work is partly supported by grants NSh-1500.2014.2, by RFBR grants 13-02-00457 (A.Mir. \& A.S.), 13-02-00478 (A.Mor.), by the joint grants 13-02-91371-ST-a, 15-51-52031-NSC-a, by 14-01-92691-Ind-a. Also we are partly supported by
the Brazilian National Counsel of Scientific and Technological Development (A.Mor.), by foundation FUNPEC-UFRN (A.Mir. \& A.S.),
and by the Quantum Topology Lab of Chelyabinsk State University (Russian Federation government grant 14.Z50.31.0020) (A.S.).

\newpage

\section*{Appendix A. Symmetrically colored HOMFLY for generalized pretzel knots}
\label{appena}

Here we present a few sample results, obtained by direct application of
evolution method of \cite{DMMSS} (see \cite{evo} for the detailed explanation).
They are the origin and justification of all the results in the main text.
Other numerous explicit formulas of this kind are too big to be included here,
still these examples are sufficient -- but only for illustrative purposes.
When some properties, like enhanced symmetry w.r.t. arbitrary permutations of
$n_i$ among parallel or antiparallel braids, are mentioned in the main text,
they were actually obtained from explicit evolution-method calculations --
not actually represented in this appendix.
After established in simpler examples, these properties were used as {\it input}
assumption in more complicated ones, thus allowing to decrease the number of
requested "initial conditions" -- still some random checks of these
assumptions were also performed at these next levels of complexity.

\subsection*{Summary, genus 2,  $n_2$ even}

In this case there is one obvious structure: for $n_2=0$  we obtain a composite knot made from two 2-strand torus knots.

Unreduced colored  HOMFLY in the lowest symmetric representations are:

\be
\chi_{[1]}H_{[1]}^{(n_1,\overline{n_2},n_3)}=
\lambda_{[2]}^{n_1+n_3}
\cdot \chi_{[2]}
\left\{1+\frac{D_2D_{-1}}{[2]}\cdot A^{n_2}\right\}
+\lambda_{[11]}^{n_1+n_3}
\cdot \chi_{[11]}
\left\{1 + \frac{D_1D_{-2}}{[2]}
\cdot A^{n_2}\right\}
+\Big(\lambda_{[2]}^{n_1}\lambda_{[11]}^{n_3}+ \lambda_{[2]}^{n_3}\lambda_{[11]}^{n_1}\Big)
\cdot [3]\chi_{22}\cdot  A^{n_2}
= \nn
\ee
\be
= \frac{1}{\chi_{[1]}^2}\left\{ H^{(n_1)}_{[1]}\overline{ H}^{(n_2)}_{[1]}H^{(n_3)}_{[1]}
+ \chi_{[2]}\chi_{[11]}\Big(\lambda_{[11]}^{n_1}-\lambda_{[2]}^{n_1}\Big)\Big(1-A^{n_2}\Big)
\Big(\lambda_{[11]}^{n_3}-\lambda_{[2]}^{n_3}\Big)\right\}
\label{H1g2}
\ee

\bigskip

\be
\chi_{[2]}H_{[2]}^{(n_1,\overline{n_2},n_3)}=
\lambda_{[4]}^{n_1+n_3}\left\{\chi_{[4]} + \frac{ \chi_{[21]}D_3D_4}{[4]^2}
\Big(D_1\cdot A^{n_2}  + \frac{D_0^2D_5}{[2]^2[3]}\cdot (qA)^{2n_2}\Big) \right\} +
 \nn \\
+\lambda_{[31]}^{n_1+n_3}\left\{\chi_{[31]} + \frac{[3] \chi_{[21]}}{[2][4]^2}
\Big( D_1\cdot U_{[31]}\cdot A^{n_2}
+ \frac{\chi_{[1]}^2D_3}{[2]} \cdot V_{[31]}\cdot(qA)^{2n_2} \Big) \right\} +
\nn\\
+\lambda_{[22]}^{n_1+n_3}\cdot \chi_{[22]}\left\{1 + \frac{D_1D_{-2}}{[2]}
\cdot A^{n_2}   + \frac{\chi_{[11]} D_3D_{-2}}{[2][3]}\cdot(qA)^{2n_2}  \right\} +
\nn \\
\nn\\
+ \Big(\lambda_{[4]}^{n_1}\lambda_{[31]}^{n_3}+\lambda_{[4]}^{n_3}\lambda_{[31]}^{n_1}\Big)
\frac{[2]\chi_{[21]}\chi_{[2]}\,D_3}{[4]^2}
\Big( A^{n_2}  + \frac{D_0D_4}{[2]^2}\cdot (qA)^{2n_2}\Big) +
\nn \\
+ \Big(\lambda_{[31]}^{n_1}\lambda_{[22]}^{n_3}+\lambda_{[31]}^{n_3}\lambda_{[22]}^{n_1}\Big)
\cdot\chi_{[22]}\chi_{[2]}\Big(A^{n_2}  + \frac{ D_3D_{-2}}{[4]}\cdot (qA)^{2n_2}\Big) +
\nn \\
+ \Big(\lambda_{[4]}^{n_1}\lambda_{[22]}^{n_3}+\lambda_{[4]}^{n_3}\lambda_{[22]}^{n_1}\Big)
\cdot \chi_{[4]}\chi_{[22]}\cdot (qA)^{2n_2}
\label{H2g2}
\ee

\bigskip

\be
\chi_{[3]}H_{[3]}^{(n_1,\overline{n_2},n_3)} \ \ = \hspace{15cm} \nn \\
\lambda_{[6]}^{n_1+n_3}\left\{\chi_{[6]} +
\frac{\chi_{[31]}D_5D_6}{[2][3]^2[4][5]^2[6]^2}
\Big(A^{n_2}\cdot [2][3]^2[4][5]D_1D_4
+(qA)^{2n_2}\cdot [3]^2[4]\chi_{[1]}^2D_3D_7
+ (q^2A)^{3n_2}\cdot [2]^2\chi_{[2]}^2D_7D_8 \Big) \right\} +
\nn \\
\lambda_{[51]}^{n_1+n_3}\left\{\chi_{[51]} + \frac{\chi_{[31]}}{[2][3]^2[4][6]^2}
\Big(A^{n_2}\cdot [2][3]^2D_1D_4\cdot U_{[51]}
+ (qA)^{2n_2}\cdot [3]\chi_{[1]}^2 D_3D_5 \cdot V_{[51]}
+   (q^2A)^{3n_2}\cdot [2]^2\chi_{[2]}^2 D_5D_6 \cdot W_{[51]} \Big) \right\} +
\nn \\
\lambda_{[42]}^{n_1+n_3}\left\{\chi_{[42]} + \frac{\chi_{[31]}\chi_{[1]}}{[2]^2[3][4][5]^2}
\Big(A^{n_2}\cdot [2][5]D_1 \cdot U_{[42]} + (qA)^{2n_2}\cdot [2]\chi_{[1]}D_3\cdot V_{[42]}
+ (q^2A)^{3n_2} \cdot  \chi_{[2]}D_{-2}D_1D_5\cdot W_{[42]} \Big) \right\} +
\nn \\
\lambda_{[33]}^{n_1+n_3}
\cdot\chi_{33}\left\{
1 + \frac{D_1D_{-2}}{[2]}\cdot A^{n_2} +  \frac{\chi_{[11]}D_3D_{-2}}{[2][3]}\cdot  (qA)^{2n_2}
   + \frac{\chi_{[22]}D_5D_{-2}}{[3][4]}\cdot(q^2A)^{3n_2}
\right\}+
\nn \\
\nn \\
 \Big(\lambda_{[6]}^{n_1}\lambda_{[51]}^{n_3}+\lambda_{[51]}^{n_3}\lambda_{[6]}^{n_1}\Big)
\frac{[2]\chi_{[31]}\chi_{[2]}D_5}{[3][4][5][6]^2}
\Big(A^{n_2}\cdot [3][4] D_4 + (qA)^{2n_2}\cdot[3]\chi_{[1]}D_3D_6
+ (q^2A)^{3n_2} \cdot\chi_{[3]}D_6D_7\Big)  +
\nn \\
 \Big(\lambda_{[6]}^{n_1}\lambda_{[42]}^{n_3}+\lambda_{[42]}^{n_3}\lambda_{[6]}^{n_1}\Big)
\frac{\chi_{[4]}\chi_{[31]}\chi_{[1]}D_5}{[2][5]^2[6]}
\Big(  (qA)^{2n_2}\cdot [2][3]
+ (q^2A)^{3n_2} D_1D_6 \Big) +
\nn \\
 \Big(\lambda_{[6]}^{n_1}\lambda_{[33]}^{n_3}+\lambda_{[33]}^{n_3}\lambda_{[6]}^{n_1}\Big)
\Big(
(q^2A)^{3n_2} \frac{\chi_{[6]}\chi_{[31]}\chi_{[2]}}{[3]^2}\Big) +
\nn \\
 \Big(\lambda_{[51]}^{n_1}\lambda_{[42]}^{n_3}+\lambda_{[42]}^{n_3}\lambda_{[51]}^{n_1}\Big)
\frac{\chi_{[31]}\chi_{[2]}\chi_{[1]}}{[2][3]^2[4][5][6]}
 \Big(A^{n_2} \cdot [2]^4[6]D_4 + (qA)^{2n_2}\cdot [2][3]D_3\cdot V_{51|42}
 + (q^2A)^{3n_2} \cdot [3]D_1D_2D_5\cdot W_{51|42}\Big) +
\nn \\
 \Big(\lambda_{[51]}^{n_1}\lambda_{[33]}^{n_3}+\lambda_{[33]}^{n_3}\lambda_{[51]}^{n_1}\Big)
\frac{[5]\chi_{[41]}\chi_{[3]}\chi_{[2]}}{[3][4]^2[6]}
\Big( (qA)^{2n_2}\cdot [6] + (q^2A)^{3n_2} \cdot D_5D_{-2}\Big) +
\nn \\
 \Big(\lambda_{[42]}^{n_1}\lambda_{[33]}^{n_3}+\lambda_{[33]}^{n_3}\lambda_{[42]}^{n_1}\Big)
\frac{\chi_{[31]}\chi_{[2]}^2}{[2][3]^2[4][5]}\Big(A^{n_2} \cdot [2][4][5]
+ (qA)^{2n_2}\cdot [2][5]D_3D_{-2} + (q^2A)^{3n_2} \cdot D_5D_2D_{-1}D_{-2}\Big)
\nn \\
\ee


Here

\be
U_{[31]} =[2]D_3D_{-2}+ (D_2-D_0)^2
\nn \\
V_{[31]} = D_3D_{-2} + D_1D_0-D_2D_{-1} = D_3D_{-2}+[2] = D_1D_0-[4]
\ee

\be
U_{[51]} = [4] D_5D_{-2} + \left( D_4-D_0 \right)^2
\nn\\
V_{[51]} = U_{[51]}  - D_2D_1[2]\{q\}^2 - D_2(D_3-D_1)[2]
\nn \\
W_{[51]} = D_5D_{-2} + [4]+[2] = D_5D_{-2} + [2][3]  = D_2D_1-[6]
\nn\\
\nn \\
U_{[42]} = D_3D_4+[2]^2D_1D_0+D_0D_{-1}-[2]^2[3][4]
\nn \\
V_{[42]} = [6]D_2D_0^2+D_0^2D_{-1} - [2]^2\Big((q^7+q^3+2q+2q^{-1}+q^{-3}+q^{-7})D_0+D_{-5}\Big)
\nn \\
W_{[42]} = [3]D_2D_1-[2][5]
\nn \\
\nn \\
V_{51|42} = D_5D_4+[2]D_2D_0+D_1D_0 - [2]^3[5]
 \nn \\
W_{51|42} = D_2D_1 - [2][5]
\ee

\bigskip

The {\it reduced} Jones polynomials ($A=q^2$):

\be
J_{[1]} = 1+ \frac{[3][4]}{[2]}\,q^{2(n_1+n_2+n_3)}
+ \nn \\
+ [3]\Big(q^{2(n_1+n_2)}+q^{2(n_1+n_3)}+q^{2(n_2+n_3)}\Big)
\ee

\be
J_{[2]} = 1+  \frac{[6]^2}{[4]^2}\,q^{2(n_1+n_2+n_3)}
+   \frac{[2][5][6][7]}{[3][4]^2}\,q^{6(n_1+n_2+n_3)}
+ \nn \\
+\frac{[2][3][5][6]}{[4]^2}
\Big(q^{6(n_1+n_2)+2n_3)}+q^{6(n_1+n_3)+2n_2)}+q^{6(n_2+n_3)+2n_1)}\Big)
+\nn \\
+\frac{[2]^2[3][5]}{[4]^2}
\Big(q^{6n_1+2(n_2+n_3)}+q^{6n_2+2(n_1+n_3)}+q^{6n_3+2(n_1+n_2)}\Big)
+\nn\\
+ [5]\Big(q^{6(n_1+n_2)}+q^{6(n_1+n_3)}+q^{6(n_2+n_3)}\Big)
+ [3]\,\Big(q^{2(n_1+n_2)}+q^{2(n_1+n_3)}+q^{2(n_2+n_3)}\Big)
\ee

\be
\!\!\!\!\!\!\!
J_{[3]} =
1+  \frac{[2][3][8]^2}{[4]^3[5]}\,q^{2(n_1+n_2+n_3)}+ \nn \\
+ \frac{[2]^3[3]^2[5][7]}{[4][6]^2}\{q\}^4 \,q^{6(n_1+n_2+n_3)}
+ \frac{[2][3][7][8][9][10]}{[4][5]^2[6]^2}\,q^{12(n_1+n_2+n_3)}
+ \nn \\
+ \frac{[2][3][7][8][9]}{[5][6]^2}\Big(q^{12(n_1+n_2)+6n_3}
+q^{12(n_1+n_3)+6n_2}+q^{12(n_2+n_3)+6n_1}\Big) + \nn \\
+\frac{[3]^2[7][8]}{[5][6]}
\Big(q^{12(n_1+n_2)+2n_3}+q^{12(n_1+n_3)+2n_2}+q^{12(n_2+n_3)+2n_1}\Big)
+\nn \\
+\frac{[2]^2[3]^2[7][8]}{[4][6]^2}\Big(q^{6(n_1+n_2)+12n_3}
+q^{6(n_1+n_3)+12n_2}+q^{6(n_2+n_3)+12n_1}\Big) + \nn \\
+\frac{[2]^2[3]^2[8]^2}{[4]^3[6]}
\Big(q^{6(n_1+n_2)+2n_3}+q^{6(n_1+n_3)+2n_2}+q^{6(n_2+n_3)+2n_1}\Big)
+ \nn \\
+ \frac{[2]^4[6]}{[4][5]}  \Big(q^{2(n_1+n_2)+6n_3}+q^{2(n_1+n_3)+6n_2}+q^{2(n_2+n_3)+6n_1}\Big)
+\nn \\ \nn \\
+ [3]\Big(q^{2(n_1+n_2)}+q^{2(n_1+n_3)}+q^{2(n_2+n_3)}\Big) +\nn \\
+ [5]\Big(q^{6(n_1+n_2)}+q^{6(n_1+n_3)}+q^{6(n_2+n_3)}\Big) +\nn \\
+ [7]\Big(q^{12(n_1+n_2)}+q^{12(n_1+n_3)}+q^{12(n_2+n_3)}\Big)
+ \nn \\
+ \frac{[2][3]^2[7]}{[5][6]}\Big(q^{12n_1+6n_2+2n_3}+q^{12n_1+6n_3+2n_2}+
q^{12n_2+6n_1+2n_3}+q^{12n_2+6n_3+2n_1}+q^{12n_3+6n_1+2n_2}+q^{12n_3+6n_2+2n_1}\Big)
\ee

\subsection*{Genus $g=3$, the first symmetric representation $[2]$
\label{Hg3r2}}

\be
\chi_{[2]}^2H^{(n_1,n_2,n_3,n_4)}_{[2]} = \lambda_{[4]}^{n_1+n_2+n_3+n_4} \cdot \left(
\frac{\chi_{[4]}^4}{\chi_{[2]}^2} + \frac{[2]}{[3]^2[4]^3}\,\chi_{[4]}\chi_{[21]}\,
\left(\frac{[3]\{Aq^4\}^2\{A\} + \{Aq^3\}^2\{Aq^2\}}{\{q\}^3}+ \frac{[3][4]\{Aq^3\}}{\{q\}}
\right)
\right)+
\ee
\be
\left( \lambda_{[4]}^{n_1+n_2+n_3}\lambda_{[22]}^{n_4} + \text{perms} \right) \left(
\frac{\chi_{[4]}^3\chi_{[22]}}{\chi_{[2]}^2} - \frac{[2]^2}{[3]^2[4]^2}\,\chi_{[4]}\chi_{[22]}\,
\frac{  \{Aq^3\}^2+\{Aq^4\}\{A\} }{  \{q\}^2  } \right)+ \nn \\
\left( \lambda_{[4]}^{n_1+n_2}\lambda_{[22]}^{n_3+n_4} + \text{perms} \right) \left(
\frac{\chi_{[4]}^2\chi_{[22]}^2}{\chi_{[2]}^2}
+ \frac{\chi_{[4]}\chi_{[32]}\chi_{[1]}}{[3]^2} \right)+\nn \\
\left( \lambda_{[4]}^{n_1}\lambda_{[22]}^{n_2+n_3+n_4} + \text{perms} \right) \left(
\frac{\chi_{[4]}\chi_{[22]}^3}{\chi_{[2]}^2}
- \frac{\chi_{[4]}\chi_{[22]}\chi_{[11]}}{[3]^2} \right)+\nn \\
\left( \lambda_{[22]}^{n_1+n_2+n_3+n_4} + \text{perms} \right) \left(
\frac{\chi_{[22]}^4}{\chi_{[2]}^2} + \frac{\chi_{[22]}(\chi_{[4]}+\chi_{[31]})}{[2]^2[3]^2}
\Big(\chi_{[11]}+[3]\Big) \right)+
\ee

\bigskip

\be
\left( \lambda_{[4]}^{n_1+n_2+n_3}\lambda_{[31]}^{n_4} + \text{perms} \right) \left(
\frac{\chi_{[4]}^3\chi_{[31]}}{\chi_{[2]}^2} - \frac{[2]}{[3][4]^3}\,\chi_{[4]}\chi_{[21]}\,
\left([2]\,\frac{q^3A^2-q^{-3}A^{-2}}{q-q^{-1}}\,\frac{\{q^3A\}}{\{q\}}
+ [3]\chi_{[1]}\right) \right) +\nn \\
\left( \lambda_{[4]}^{n_1+n_2}\lambda_{[31]}^{n_3+n_4} + \text{perms} \right) \left(
\frac{\chi_{[4]}^2\chi_{[31]}^2}{\chi_{[2]}^2} + \frac{[2]}{[4]^3}\,\chi_{[4]}\chi_{[21]}\,
\left(\frac{q^4A^3-q^{-4}A^{-3}}{q-q^{-1}}+[2]\frac{\{qA\}}{\{q\}}\right) \right)+ \nn \\
\left( \lambda_{[4]}^{n_1}\lambda_{[31]}^{n_2+n_3+n_4} + \text{perms} \right) \left(
\frac{\chi_{[4]}\chi_{[31]}^3}{\chi_{[2]}^2} - \frac{[3]}{[4]^3}\,\chi_{[4]}\chi_{[21]}\,
\left( (q^2A^3+q^{-2}A^{-3}) + 2\frac{\{qA\}}{\{q\}}
\right)
\right)+
\ee
\be
\lambda_{[31]}^{n_1+n_2+n_3+n_4} \left(
\frac{\chi_{[31]}^4}{\chi_{[2]}^2} + \frac{1}{[4]^3}\,\chi_{[31]}\chi_{[2]}\,
\left( ( q^3A^4 {+} {1\over q^{3}A^{4}} )  {+}  \frac{(4q^3{-} 3q{+} {2\over q} {+} {1\over q^3} )A^2 {-} 2[4]
{+}( {4\over q^3} {-} {3\over q} {+} 2q {+} q^3 ) {1\over A^2} }{(q-q^{-1})^2}
\right)
\right)+
\ee
\be
\left( \lambda_{[31]}^{n_1+n_2+n3}\lambda_{[22]}^{n_4} + \text{perms} \right) \left(
\frac{\chi_{[22]}\chi_{[31]}^3}{\chi_{[2]}^2} + \frac{[3]}{[4]^3}\,\chi_{[22]}\chi_{[3]}\,
\left( (q^2A^3+q^{-2}A^{-3}) - 2\frac{\{A/q\}}{\{q\}}-(q^3+q^{-3})\chi_{[1]}\right) \right)+ \nn \\
\left( \lambda_{[31]}^{n_1+n_2}\lambda_{[22]}^{n_3+n_4} + \text{perms} \right) \left(
\frac{\chi_{[22]}^2\chi_{[31]}^2}{\chi_{[2]}^2} + \frac{1}{[2][4]^2}\,\chi_{[22]}\chi_{[3]}
\chi_{[1]}\,\Big((qA^2+q^{-1}A^{-2}) + [2][3]\Big)  \right) +\nn \\
\left( \lambda_{[31]}^{n_1}\lambda_{[22]}^{n_2+n_3+n_4} + \text{perms} \right) \left(
\frac{\chi_{[22]}^3\chi_{[31]}}{\chi_{[2]}^2} + \frac{1}{[2]^2[3][4]}\,\chi_{22]}\chi_{[3]}
\chi_{[1]}\,\Big((A+A^{-1})\chi_{[1]}-[2][3]\Big) \right)+
\ee
\be
\left( \lambda_{[4]}^{n_1+n_2}\lambda_{[31]}^{n_3}\lambda_{[22]}^{n_4} + \text{perms} \right) \left(
\frac{\chi_{[4]}^2\chi_{[31]}\chi_{[22]}}{\chi_{[2]}^2}
+ \frac{[2]^2}{[3][4]^2}\,\chi_{[4]}\chi_{[22]}\,\frac{q^3A^2-q^{-3}A^{-2}}{q-q^{-1}}  \right) +\nn \\
\left( \lambda_{[4]}^{n_1}\lambda_{[31]}^{n_2+n_3}\lambda_{[22]}^{n_4} + \text{perms} \right) \left(
\frac{\chi_{[4]}\chi_{[31]}^2\chi_{[22]}}{\chi_{[2]}^2}
- \frac{[2]}{[4]^2}\,\chi_{[4]}\chi_{[22]}\,(qA^2+q^{-1}A^{-2})  \right) +\nn\\
 \nn \\
\left( \lambda_{[4]}^{n_1}\lambda_{[31]}^{n_2}\lambda_{[22]}^{n_3+n_4} + \text{perms} \right) \left(
\frac{\chi_{[4]}\chi_{[31]}\chi_{[22]}^2}{\chi_{[2]}^2}
- \frac{1}{[3][4]}\,\chi_{[4]}\chi_{[22]}\chi_{[1]}\,(A+A^{-1}) \right)
\ee

The result (\ref{poldeco1}) of the present paper is that this long
and strangely-looking expression is nothing else but
\be
H^{(n_1,n_2,n_3,n_4)}_{[2]} =
\sum_{k=0}^3 \bar\Delta_k \cdot \prod_{i=1}^4 \left(\sum_{m=0}^3 a_{km}\lambda_m^{n_i}\right)
\ \stackrel{(\ref{A3})}{=} \
\prod_{i=1}^4
\left(\frac{D_0D_{-1}}{[2][3]} + \frac{D_2D_{-1}}{[4]}\cdot (-q^2)^{n_i} +
\frac{D_3D_2}{[3][4]}\cdot q^{6n_i}\right) +
\nn\\
+D_1D_{-1}\cdot  \prod_{i=1}^4\left(\frac{D_0}{[2][3]} + \frac{D_2-D_0}{[4]}\cdot (-q^2)^{n_i}
- \frac{[2]D_3}{[3][4]} \cdot q^{6n_i}\right)
+\frac{D_3D_0^2D_{-1}}{[2]^2}\cdot  \prod_{i=1}^4\left(\frac{1}{[3]} - \frac{[2]}{[4]}\cdot (-q^2)^{n_i}
+ \frac{[2] }{[3][4]} \cdot q^{6n_i}\right)
\nn
\ee
which is not only shorter, but also a much better structured expression,
moreover, generalizable to arbitrary genus and representation.

\newpage

\section*{Appendix B. List of coefficients $a_{km}$, $\bar a_{km}$ and $\bar{\bar{a}}_{km}$}
\label{appenb}

We list in this Appendix both the coefficients of all three matrices ${\cal A}$, $\bar{\cal A}$ and ${\cal \bar{\bar A}}$ and the two Racah matrices corresponding to ${\cal A}$, $\bar{\cal A}$, since the third Racah matrix is obtained from that for ${\cal A}$ just by transposing.

\subsection*{Coefficients $a_{km}$}

Coefficients $a_{km}$ entering contributions of the parallel braid:
\be
 {\cal A}_{[1]} = \frac{1}{\chi_{[1]}}\left(\begin{array}{cc}
\chi_{[11]} & \chi_{[2]}\\ \\ \cfrac{\chi_{[1]}}{[2]}& -\cfrac{\chi_{[1]}}{[2]}
\end{array}\right) =
\frac{1}{[2]}\left(\begin{array}{cc}
D_{-1} & D_1 \\ \\ 1 & -1 \end{array}\right)
\nn
\ee

\be
{\cal A}_{[2]} =
\frac{1}{\chi_{[2]}}
\left(\begin{array}{ccc}
 {\chi_{[22]}}  &  {\chi_{[31]}}  &  {\chi_{[4]}}  \\
&&\\
\cfrac{\chi_{[2]}\cdot D_0}{[2][3]} & \cfrac{\chi_{[2]}\cdot(D_2-D_0)}{[4]}
&  -\cfrac{[2]\,\chi_{[2]}\cdot D_3}{[3][4]}  \\
&&\\
\cfrac{\chi_{[2]}}{[3]} &  -\cfrac{[2]\,\chi_{[2]}\cdot}{[4]}  &
\cfrac{[2]\,\chi_{[2]}\cdot}{[3][4]}
\end{array}\right)
= \frac{1}{[3]}
 \left(\begin{array}{ccc}
\cfrac{1}{[2]}\,D_0D_{-1} & \cfrac{[3]}{[4]}\,D_2D_{-1} & \cfrac{1}{[4]}\,D_3D_2 \\
&&\\
\cfrac{1}{[2]}\,D_0 & \cfrac{[3]}{[4]}\,(D_2-D_0)
&  -\cfrac{[2]}{[4]}\,D_3  \\
&&\\
1 &  -\cfrac{[2][3]}{[4]}  &  \cfrac{[2]}{[4]}
\end{array}\right)
\nn
\ee

\be
{\cal A}_{[3]} =
\frac{1}{[4]}\left(\begin{array}{cccc}
\cfrac{1}{[2][3]}\,D_1D_0D_{-1} &\cfrac{[3]}{[2][5]}\,D_3D_0D_{-1} &\cfrac{1}{[6]}\,D_4D_3D_{-1}&
\cfrac{1}{[5][6]}\,D_5D_4D_3 \\
&&& \\
\cfrac{1}{[2][3]}\,D_1D_0& \cfrac{1}{[2][5]}\Big([2]D_4-D_{-1}\Big)D_0
& \ \cfrac{1}{[3][6]}\,D_4\Big(D_5-[2]^2D_{-1}\Big)
& -\cfrac{[3]}{[5][6]}\,D_5D_4 \\
&&& \\
\cfrac{1}{[3]}\,D_1 &\cfrac{1}{[5]}\Big(D_5-[2]D_0\Big)
&\cfrac{[2]}{[3][6]}\Big(-[2]^2D_5+D_{-1}\Big)
& \cfrac{[2][3]}{[5][6]}\,D_5 \\
&&&\\
1 &-\cfrac{[3]^2}{[5]}&\cfrac{[2][3]}{[6]}& -\cfrac{[2][3]}{[5][6]}
\end{array} \right)
\label{A3}
\ee

\bigskip


\centerline{
{\footnotesize
$ \ {\cal A}_{[4]} = \frac{1}{[5]}
\left(\begin{array}{ccccc}
\cfrac{D_2D_1D_0D_{-1}}{[2][3][4]} & \cfrac{1}{[2][6]}\,D_4\,D_1D_0D_{-1}
&\frac{[5]}{[2][6][7]}\,D_5D_4\,D_0D_{-1} & \cfrac{1}{[6][8]}\,D_6D_5D_4\,D_{-1}
&\cfrac{D_7D_6D_5D_4}{[6][7][8]} \\
&&&&\\
\!\!\!\frac{D_2D_1D_0}{[2][3][4]}
&\!\!\!\!\!\!\!\!\cfrac{D_1D_0}{[2][4][6]}\,\Big([3]D_5 {-} D_{-1}\Big)
&\cfrac{[5]}{[4][6][7]}\cdot D_5\Big(D_6 {-} [2] D_{-1}\Big)D_0
& \!\!\!\!\!\!\!\cfrac{D_6D_5}{[4][6][8]}\Big(D_7 {-} [3]^2D_{-1}\Big)
& -\cfrac{[4]}{[6][7][8]} D_7D_6D_5 \\
&&&&\\
\frac{1}{[3][4]}\,D_2D_1
&\!\!\!\!\!\!\!\!\!\!\!\!\cfrac{[2]}{[4][6]}\Big(D_6-D_0\Big)D_1
&\!\!\!\!\!\!\!\!\!\!
\cfrac{[2][5]}{[3][4][6][7]}\Big(D_7D_6 {-} [2]^3D_6D_0 {+} D_0D_{{-}1}\Big)
&\frac{[2]^2[3]}{[4][6][8]}\,D_6\Big({-}D_7 {+} D_{-1}\Big)
&  \cfrac{[3][4]}{[6][7][8]}\, D_7D_6 \\
&&&&\\
\cfrac{1}{[4]}\,D_2
&\cfrac{[3]}{[4][6]}\Big(D_7-[3]D_1\Big)
&\cfrac{[2][3][5]}{[4][6][7]}\Big({-}[2]D_7 {+} D_0\Big)
&\cfrac{[2][3]}{[4][6][8]}\Big([3]^2D_7 {-} D_{-1}\Big)
&-\cfrac{[2][3][4]}{[6][7][8]}\,D_7\\
&&&&\\
1 &-\cfrac{[3][4]}{[6]}&\cfrac{[3][4][5]}{[6][7]}&-\cfrac{[2][3][4]}{[6][8]}
&  \cfrac{[2][3][4]}{[6][7][8]}
\end{array}\right)
$
}}

\bigskip

The Racah matrix associated with $a_{km}$ are
\be
 S_{[1]} =  \frac{1}{\sqrt{[2]\,D_0}}
\left(\begin{array}{cc} \sqrt{D_{-1}} & \sqrt{D_1} \\
\sqrt{D_1}& -\sqrt{D_{-1}} \end{array}\right)
\nn
\ee

\be
S_{[2]} = \frac{1}{[3]}
 \left(\begin{array}{ccc}
\sqrt{\dfrac{1}{\Delta_0}}  \cdot \cfrac{1}{[2]}\,D_0D_{-1} & \sqrt{\dfrac{1}{\Delta_1}} \cdot \cfrac{[3]}{[4]}\,D_2D_{-1} & \sqrt{\dfrac{1}{\Delta_2}} \cdot \cfrac{1}{[4]}\,D_3D_2 \\
&&\\
\sqrt{\dfrac{\bar \Delta_1}{\Delta_0}} \cdot \cfrac{1}{[2]}\,D_0 & \sqrt{\dfrac{\bar \Delta_1}{\Delta_1}} \cdot \cfrac{[3]}{[4]}\,(D_2-D_0)
&  -\sqrt{\dfrac{\bar \Delta_1}{\Delta_2}} \cdot \cfrac{[2]}{[4]}\,D_3  \\
&&\\
\sqrt{\dfrac{\bar \Delta_2}{\Delta_0}} &  -\sqrt{\dfrac{\bar \Delta_2}{\Delta_1}} \cdot \cfrac{[2][3]}{[4]}  &  \sqrt{\dfrac{\bar \Delta_2}{\Delta_2}} \cdot \cfrac{[2]}{[4]}
\end{array}\right)
\ee

\subsection*{Coefficients $\bar a_{km}$}

Coefficients $\bar a_{km}$ entering contributions of the antiparallel braid with even crossings:

\be
\bar{\cal A}_1
= \frac{1}{D_0}\left(\begin{array}{cc}  1 & D_1 D_{-1} \\
&\\ 1 & -1
\end{array}\right)
=  \frac{1}{\chi_{[1]}}\left(\begin{array}{cc}  1 &  \bar \Delta_1 \\
&\\ 1 & -1
\end{array}\right),
\nn \\ \nn \\
\nn \\ \nn \\
\bar{\cal A}_2=  \frac{1}{\chi_{[2]}}\left(\begin{array}{ccc}  1 &  \bar \Delta_1 & \bar \Delta_2 \\
&&\\
1& \cfrac{D_1}{[2]D_2}\,\Big(D_3D_{-1}-1\Big) & -\cfrac{D_0^2D_3}{[2]D_2}\\
&&\\
1&  -[2]\cfrac{D_1}{D_2} & \cfrac{D_0}{D_2}
\end{array}\right),
\nn \\ \nn \\
\nn \\ \nn \\
\bar{\cal A}_3=  \frac{1}{\chi_{[3]}}\left(\begin{array}{cccc}  1 &  \bar \Delta_1&\bar \Delta_2&\bar \Delta_3\\
&&&\\
1 &\cfrac{D_1}{[3]\,D_3}\Big([2]\,D_3D_0-[3]^2\Big)&\cfrac{D_0^2}{[2]^2[3]}\Big(D_4D_0-[3]^2\Big)
&-\cfrac{D_5D_1^2D_0^2}{[2]^2[3]\,D_3}\\
&&&\\
1&\cfrac{D_1}{[3]\,D_3}\Big(D_4D_0-[3]^2\Big)&-\cfrac{D_0}{[3]\,D_4}\Big([2]\,D_4D_1-[3]^2\Big)
&\cfrac{D_5D_1^2D_0}{[3]\,D_4D_3}\\
&&&\\
1&-\cfrac{[3]\,D_1}{D_3}&\cfrac{[3]\,D_0}{D_4}&-\cfrac{D_1D_0}{D_4D_3} \\
\end{array}\right),
\nn \\ \nn \\
\label{antiA}
\ee
\bigskip

\centerline{{\footnotesize
$
\bar{\cal A}_4 =  \frac{1}{\chi_{[4]}}\left(\begin{array}{ccccc}
1 &  \bar \Delta_1 &  \bar \Delta_2&\bar \Delta_3&\bar \Delta_4\\
&&&&\\
1 &\cfrac{D_1}{[4]\,D_4}
\Big([3]\,D_4D_0-[4]^2\Big)&\cfrac{D_3D_0^2}{[2][4]\,D_4}\Big(D_5D_0 {-} \frac{[4]^2}{[2]}\Big)
&\cfrac{D_5D_1^2D_0^2}{[2]^2[3]^2[4]\,D_4}\Big(D_6D_0 {-} [4]^2\Big)
&\!\!\!\!\!\!\!\!-\cfrac{D_7D_2^2D_1^2D_0^2}{[2]^2[3]^2[4]\,D_4} \\
&&&&\\
1 &\cfrac{[2]\,D_1}{[4]\,D_4}\Big(D_5D_0 {-} \frac{[4]^2}{[2]}\Big)
&\cfrac{D_3D_0}{[2][3][4]\,D_5D_4}\Big(D_6^2D_0^2 {-} \big([7] {+} [2]^4\big)D_6D_0 {+} [2]^2\Big)
&\!\!\!\!\!\!\!\!\!\! -\cfrac{D_1^2D_0}{[3][4]\,D_4}\Big(D_7D_0 {-} [2]\Big)
& \cfrac{D_7D_2^2D_1^2D_0}{[2][3][4]\,D_5D_4}\\
&&&&\\
1 &\cfrac{D_1}{[4]\,D_4}\Big(D_6D_0-[4]^2\Big)
&-\cfrac{[3]\,D_3D_0}{[4]\,D_5D_4}\Big(D_7D_0 {-} [2]\Big)
&\!\!\!\!\!\!\!\!\!\!\cfrac{D_1D_0}{[4]\,D_6D_4}\Big([3]D_8D_0 {+} [2][8] {+} \frac{[2][6]}{[3]}\Big)
& -\cfrac{D_7D_2^2D_1D_0}{[4]\,D_6D_5D_4}\\
&&&&\\
1 &-\cfrac{[4]\,D_1}{D_4}&\cfrac{[3][4]\,D_3D_0}{[2]\,D_5D_4}&-\cfrac{[4]\,D_1D_0}{D_6D_4}
& \cfrac{D_2D_1D_0}{D_6D_5D_4}
\end{array}\right)
$
}}

 \bigskip

The Racah matrix associated with $\bar a_{km}$ are

\be
\nn \\
\bar S_1
=  \frac{1}{\chi_{[1]}}\left(\begin{array}{cc}  1 &  \sqrt{\bar \Delta_1} \\
&\\ \sqrt{\bar \Delta_1} & -1
\end{array}\right),
\nn \\ \nn \\
\nn \\ \nn \\
\bar S_2=  \frac{1}{\chi_{[2]}}\left(\begin{array}{ccc}  1 &  \sqrt{\bar \Delta_1} & \sqrt{\bar \Delta_2} \\
&&\\
\sqrt{\bar \Delta_1}& \cfrac{D_1}{[2]D_2}\,\Big(D_3D_{-1}-1\Big) & -\cfrac{D_0}{D_2}\sqrt{D_3D_1}\\
&&\\
\sqrt{\bar \Delta_2}&  -\cfrac{D_0}{D_2}\sqrt{D_3D_1} & \cfrac{D_0}{D_2}
\end{array}\right)
\ee

\subsection*{Coefficients $\bar {\bar a}_{km}$}

Coefficients $\bar {\bar a}_{km}$ entering contributions of the antiparallel braid with odd crossings:

\be
\overline{\overline{{\cal A}}}_{[1]} = \frac{1}{D_0}\left(\begin{array}{cc}
1 & D_1 \\ 1 & - D_{-1} \end{array}\right)
\nn
\ee
\be
\overline{\overline{{\cal A}}}_{[2]}
= \frac{1}{[3]}
 \left(\begin{array}{ccc}
\cfrac{1}{[2]\chi_{[22]}}\,D_0D_{-1}
& \cfrac{\Delta_1}{[2]\chi_{[22]}}\,D_0
& \cfrac{\Delta_2}{\chi_{[22]}} \\
&&\\
\cfrac{[3]}{[4]\chi_{[31]}}\,D_2D_{-1}
& \cfrac{[3]\Delta_1}{[4]\chi_{[31]}}\,(D_2-D_0)
&   -\cfrac{[2][3]\Delta_2}{[4]\chi_{[31]}}  \\
&&\\
\cfrac{1}{[4]\chi_{[4]}}\,D_3D_2
& -\cfrac{[2]\Delta_1}{[4]\chi_{[4]}}\,D_3  &  \cfrac{[2]\Delta_2}{[4]\chi_{[4]}}
\end{array}\right)
= \frac{1}{D_0D_1D_2}\left(\begin{array}{ccc}
\l[2]\,D_2 & \l[2]\,D_2D_1 & {D_3D_2D_0} \\ \\
\l[2]\,D_2 & \l[2]\,D_1\Big(D_2-D_0\Big) & -D_3D_0^2\\ \\
\l[2]\,D_2 & -[2]^2D_1D_{-1}& D_0^2D_{-1}
\end{array}\right)
\nn
\ee

\be
\overline{\overline{{\cal A}}}_{[3]} =
\frac{1}{[4]}\left(\begin{array}{cccc}
\cfrac{1}{[2][3]\chi_{[33]}}\,D_1D_0D_{-1}
&  \cfrac{\Delta_1}{[2][3]\chi_{[33]}}\,D_1D_0
&  \cfrac{\Delta_2}{[3]\chi_{[33]}}\,D_1
&\cfrac{\Delta_3}{\chi_{[33]}} \\
&&& \\
\cfrac{[3]}{[2][5]\chi_{[42]}}\,D_3D_0D_{-1}
& \cfrac{\Delta_1}{[2][5]\chi_{[42]}}\Big([2]D_4-D_{-1}\Big)D_0
&\cfrac{\Delta_2}{[5]\chi_{[42]}}\Big(D_5-[2]D_0\Big)
& -\cfrac{[3]^2\Delta_3}{[5]\chi_{[42]}} \\
&&& \\
\cfrac{1}{[6]\chi_{[51]}}\,D_4D_3D_{-1}
& \ \cfrac{\Delta_1}{[3][6]\chi_{[51]}}\,D_4\Big(D_5-[2]^2D_{-1}\Big)
& \cfrac{[2]\Delta_2}{[3][6]\chi_{[51]}}\Big(-[2]^2D_5+D_{-1}\Big)
&   \cfrac{[2][3]\Delta_3}{[6]\chi_{[51]}}  \\
&&&\\
\cfrac{1}{[5][6]\chi_{[6]}}\,D_5D_4D_3
&  -\cfrac{[3]\Delta_1}{[5][6]\chi_{[6]}}\,D_5D_4
& \cfrac{[2][3]\Delta_2}{[5][6]\chi_{[6]}}\,D_5
& -\cfrac{[2][3]\Delta_3}{[5][6]\chi_{[6]}}
\end{array} \right)
= \nn \\ \nn \\ \nn \\ \nn \\ =
\frac{1}{D_4D_3D_2D_1D_0}\left(\begin{array}{cccc}
\l[2][3]\,D_4D_3
& \l[2][3]\,D_4D_3D_1
&  \l[3]\,D_4D_3^2D_0
&D_5D_4D_3D_1D_0 \\
&&& \\
\l[2][3]\,D_3D_4
& \l[2]\,D_4D_1\Big([2]D_4-D_{-1}\Big)
&D_4D_3D_0\Big(D_5-[2]D_0\Big)
& -D_5D_4D_1^2D_0 \\
&&& \\
 \l[2][3]\,D_4D_3
&  \l[2]\,D_4D_1\Big(D_5-[2]^2D_{-1}\Big)
& D_3D_0^2\Big(-[2]^2D_5+D_{-1}\Big)
&   D_5D_1^2D_0^2 \\
&&&\\
 \l[2][3]\,D_4D_3
&  -[2][3]^2D_4D_1D_{-1}
&  \l[3]^2D_3D_0^2D_{-1}
& -D_1^2D_0^2D_{-1}
\end{array} \right)
\label{bbA3}
\ee

\end{document}